\documentclass[journal]{IEEEtran}
\usepackage{amsmath,amsfonts}
\usepackage{algorithmic}
\usepackage{algorithm}
\usepackage{array}
\usepackage[caption=false,font=normalsize,labelfont=sf,textfont=sf]{subfig}
\usepackage{textcomp}
\usepackage{stfloats}
\usepackage{url}
\usepackage{verbatim}
\usepackage{graphicx}
\usepackage[compatibility=false]{caption}
\usepackage{amssymb}
\usepackage{xcolor}
\usepackage[normalem]{ulem}
\usepackage{subcaption}
\usepackage{siunitx}
\usepackage{cite} 
\def\BibTeX{{\rm B\kern-.05em{\sc i\kern-.025em b}\kern-.08em
    T\kern-.1667em\lower.7ex\hbox{E}\kern-.125emX}}
\usepackage{balance}

\begin{document}
\title{A Novel Stripe-based RIS Optimization for UAV Communications and Sensing in Low-Altitude Wireless Networks}

\author{
Burak~Ahmet~Çelebi,         
Sefa~Kayraklik,
Onur~Salan,~\IEEEmembership{Graduate Student Member,~IEEE},
İbrahim~Hökelek,~\IEEEmembership{Member,~IEEE,}
Ali~Emre~Pusane,
Ali~Görçin,~\IEEEmembership{Senior Member,~IEEE,
}   
        \thanks{B. A. Çelebi, S. Kayraklık, O. Salan, İ. Hökelek, and A. Görçin are with the Communications and Signal Processing Research (HİSAR) Lab., TÜBİTAK-BİLGEM, Kocaeli, Turkiye. E-mail: \{burak.celebi, sefa.kayraklik, onur.salan, ibrahim.hokelek, ali.gorcin\}@tubitak.gov.tr}
        
        \thanks{B. A. Çelebi and A. E. Pusane are with the Department of Electrical and Electronics Engineering, Boğaziçi University, Sariyer, Istanbul, Turkiye. E-mail: ali.pusane@bogazici.edu.tr}
        
        \thanks{S. Kayraklık is with the Department of Electrical and Electronics Engineering, Koç University, Sariyer, Istanbul, Turkiye.}
        
        \thanks{O. Salan, and A. Görçin are with the Department of Electrical and Electronics Engineering, İstanbul Technical University, Sariyer, Istanbul, Turkiye.}
        \vspace{-10pt}
}

\maketitle

\begin{abstract}
Low-altitude wireless networks (LAWN) envision a reconfigurable 3D network capable of supporting mission-critical aerial operations. This paper presents a reconfigurable intelligent surface (RIS)-assisted LAWN to establish a reliable communication with an unmanned aerial vehicle (UAV) across varying wireless channel conditions and signal blockages. A low complexity stripe-based RIS phase shift optimization framework is proposed to simultaneously enhance communication reliability and provide passive sensing capability for UAV tracking under 3D mobility. Unlike high-complexity optimization approaches, the proposed method leverages the inherent structural phase-gradient of the RIS adjacent elements to significantly reduce the search space for calculating and updating the RIS configuration as the UAV moves. The analysis and simulation results demonstrate that the proposed framework outperforms conventional benchmarks in convergence speed and computational efficiency, while maintaining robust, high signal-to-noise-ratio (SNR) connectivity even in the presence of phase estimation errors and low SNR regimes. In addition, the measurement experiments using a real RIS prototype in an outdoor campus environment are performed to demonstrate the practical viability of the proposed approach.

\end{abstract}


\section{Introduction}
The emergence of low-altitude wireless networks (LAWN) introduces a new digital infrastructure paradigm characterized by a multifunctional, service-aware, and dynamically reconfigurable three-dimensional (3D) networking architecture \cite{yuan2025ground}. By integrating communication, sensing, control, and distributed computing across aerial and terrestrial nodes, LAWN empowers unmanned aerial vehicles (UAVs) to execute intelligent, mission-driven operations across domains ranging from aerial logistics to public safety  \cite{8918497}. However, the highly dynamic nature of these operations necessitates more than simple aerial connectivity, requiring a mechanism to cope with rapidly fluctuating wireless channels while ensuring the reliability and low latency for cooperative control. In this context, reconfigurable intelligent surfaces (RISs) have emerged as a flexible physical layer enabler to maintain the 3D network by controlling the wireless propagation environment, thereby enhancing the coverage and reliability of UAV-based communication systems across both aerial and terrestrial nodes \cite{9124704}.

Recent literature has extensively explored integrating RISs into LAWNs to enhance aerial connectivity \cite{8959174,9120632,9277627,9684973,9322325}. A preliminary study focuses on maximizing the achievable rate, where a successive convex approximation is employed to jointly optimize a UAV trajectory and RIS phase shifts \cite{8959174}. From a coverage perspective, the impact of RIS deployment on cellular communication supporting UAVs is analysed in terms of UAV height and RIS parameters, including size, altitude, and distance from the base station \cite{9120632}. Moving beyond convex optimization, a deep reinforcement learning (DRL)-based framework is proposed to minimize energy consumption in non-orthogonal multiple access-based networks by jointly determining RIS phase shifts, UAV trajectory, and power allocation policies \cite{9277627}. Another DRL-based scheme is implemented to maximize the energy efficiency by considering both the UAV's 3D trajectory and the RIS phase shifts \cite{9684973}. Furthermore, the sum data rate maximization in an orthogonal frequency division multiple access system is studied to determine the UAV's trajectory, RIS scheduling, and resource allocation through an alternating optimization method \cite{9322325}.

Beyond basic connectivity provisioning, recent studies have extended RIS-aided LAWN frameworks to support diverse mission-critical applications \cite{9400768,9416239,10289638,11105378}. For example, a symbiotic radio system is proposed in which a UAV assists the RIS in conveying environmental information. The objective is to minimize the RIS bit error rate through the joint optimization of the UAV trajectory, RIS scheduling, and phase-shift configuration \cite{9400768}. As another application, an RIS-assisted UAV physical layer security system is investigated to optimize the worst-case secrecy rate through the joint design of the UAV’s mobility, the RIS’s phase shifts, and the transmitter power \cite{9416239}. Furthermore, a wireless-powered communication system for RIS-assisted UAV networks is studied, enabling multiple passive Internet of Things devices to gather energy from transmitted signals \cite{10289638}. Additionally, network management is realized through digital twin architectures, which enable real-time resource allocation for RIS-aided next-generation wireless communication systems \cite{11105378}.

More recently, integrated sensing and communication (ISAC) systems have attracted considerable attention in RIS-empowered LAWNs due to their ability to simultaneously provide information about the physical environment while maintaining the communication link. Regarding the physical layer design of a UAV-enabled ISAC system, integrated waveform optimization is proposed to minimize multi-user interference by balancing beam pattern mismatches under peak-to-average power ratio constraints \cite{10892027}. To effectively handle the complexity of such systems, DRL-based techniques have been widely employed \cite{11079608,11205853}. For instance, a multi-agent DRL framework is developed for an RIS-assisted UAV-enabled ISAC system with dual-functional antennas at the aerial transceiver to dynamically balance sensing and communication resources \cite{11079608}. In addition, a mixture-based proximal policy optimization approach is employed to maximize sensing performance in a DRL-driven transmission scheme for a hybrid RIS-assisted cellular ISAC architecture within a UAV-enabled LAWN \cite{11205853}. Physical layer security is another important research area in RIS-assisted ISAC-LAWN \cite{11207524, yigit2025dual}. For example, a recent study focuses on optimizing worst-case secrecy rates in the presence of an eavesdropper and a untrusted target by co-designing UAV trajectories, RIS phase shifts, and transmit/receive beamforming \cite{11207524}. In another study, a dual UAV-mounted RIS-aided ISAC architecture is introduced to mitigate malicious interference while satisfying sensing constraints \cite{yigit2025dual}. A comprehensive survey on the characteristics and challenges of RIS-aided LAWN with particular emphasis on UAV-ISAC networks can be found in \cite{11361143}.

To bridge the gap between theory and practical RIS-assisted LAWN deployment, recent efforts have begun to incorporate realistic hardware and environmental constraints, UAV mobility, and practical RIS models. Regarding hardware imperfections, \cite{10666751} incorporates transceiver impairments and practical phase shift models into spectral efficiency optimization. Practical RIS optimization methods, including element-by-element iterative adjustment, grouping neighboring elements under a common phase configuration, and predefined codebooks for rapid beam selection, are implemented and tested in real-world measurement setups to enhance signal coverage \cite{10278759} and provide physical layer security \cite{10423876}. On the mobility front, the energy efficiency of an RIS-assisted LAWN is studied under UAV jitters and imperfect hardware constraints, with the joint optimization of RIS phase shifts, UAV beamforming, and UAV trajectory \cite{10075533}. The effect of hardware impairments on 3D UAV tracking and localization in an RIS-assisted system is analyzed in \cite{10551819}. More recently, RIS-assisted beam tracing methods for LAWNs are proposed to enhance the stability of dynamic communication links \cite{11363050}.

Against this background, practical RIS optimization in UAV-based LAWNs has not been sufficiently explored to maintain reliable communication links while simultaneously enabling UAV tracking. In this paper, a stripe-based RIS phase shift optimization strategy for both stationary and mobile UAVs is proposed to improve signal quality at the UAV. Under the discrete phase shift adjustment, the proposed RIS optimization method reduces the computational complexity from linear (iterative) to sublinear (stripe-based) with respect to the number of RIS elements. In addition, the stripe-based RIS optimization is more robust across different signal-to-noise ratio (SNR) regimes, whereas the iterative algorithm's performance is significantly degraded at low SNR, as the improvement from a single RIS element remains below the noise floor. Additionally, the RIS is leveraged to enable sensing functionality by estimating the UAV direction through tracking of the applied phase-shift configurations. The average SNR loss results obtained from both analytical derivations and Monte Carlo simulations exhibit close agreement, validating the phase-estimation error analysis of the stripe-based method. Furthermore, measurement experiments using a real RIS prototype and software-defined radios (SDRs) in an outdoor campus environment demonstrate the practical viability of the proposed approach. The contributions of this paper are summarized as follows:
\begin{itemize}
    \item A system model for an RIS-assisted LAWN is developed to establish a reliable communication channel from a base station to a UAV in addition to providing a complementary passive sensing capability for UAV tracking, explicitly incorporating discrete RIS phase-shift optimization and realistic 3D UAV mobility. 
    \item A low-complexity stripe-based RIS optimization algorithm for a stationary UAV is developed. A mathematical analysis is presented to characterize the system performance under structured phase-gradient estimation errors.
    \item The stripe-based optimization framework is extended with Kalman-aided and extremum seeking control (ESC)-aided novel tracking methodologies, where the UAV’s direction is estimated through the RIS phase-gradient between adjacent elements. A mathematical transformation that converts RIS phase-shift optimization into a continuous UAV tracking mechanism is derived, effectively providing the system with angular localization as a complementary sensing capability.
    \item Comprehensive analysis and simulation results for the SNR performance across varying UAV distances, the algorithmic runtime, the effects of phase estimation error, and the UAV tracking accuracy are reported. Furthermore, the proposed framework's practical feasibility is demonstrated through outdoor measurement experiments using a real RIS prototype and SDRs.
\end{itemize}

The organization of this paper is as follows. Section II presents the system model along with the problem formulation. Section III describes the stripe-based RIS phase shift optimization for a stationary UAV under phase estimation error. Section IV considers a mobile UAV scenario, where the stripe-based approach is extended for UAV tracking and angular localization. Section V provides the simulation results and discussion, followed by the conclusion in Section VI.

\textit{Notation:} Throughout this paper, scalars are denoted by standard italic letters (e.g., $x$), whereas vectors and matrices are represented by boldface lowercase (e.g., $\mathbf{x}$) and boldface uppercase (e.g., $\mathbf{X}$) letters, respectively. The calligraphic uppercase letters (e.g., $\mathcal{X}$) are used to indicate sets. $\mathrm{diag}(\mathbf{\mathbf{x}})$ describes a diagonal matrix whose diagonal elements are the elements of the vector $\mathbf{x}$. The operators $|\cdot|$ and $\|\cdot\|$ denote the absolute value of a scalar and the Euclidean ($\ell_2$) norm of a vector, respectively. Finally, the statistical expectation of a random variable is denoted by $\mathbb{E}[\cdot]$. 

\section{System Model} \label{SystemModel}

\begin{figure}[t]
    \centering
    \includegraphics[width=\linewidth]{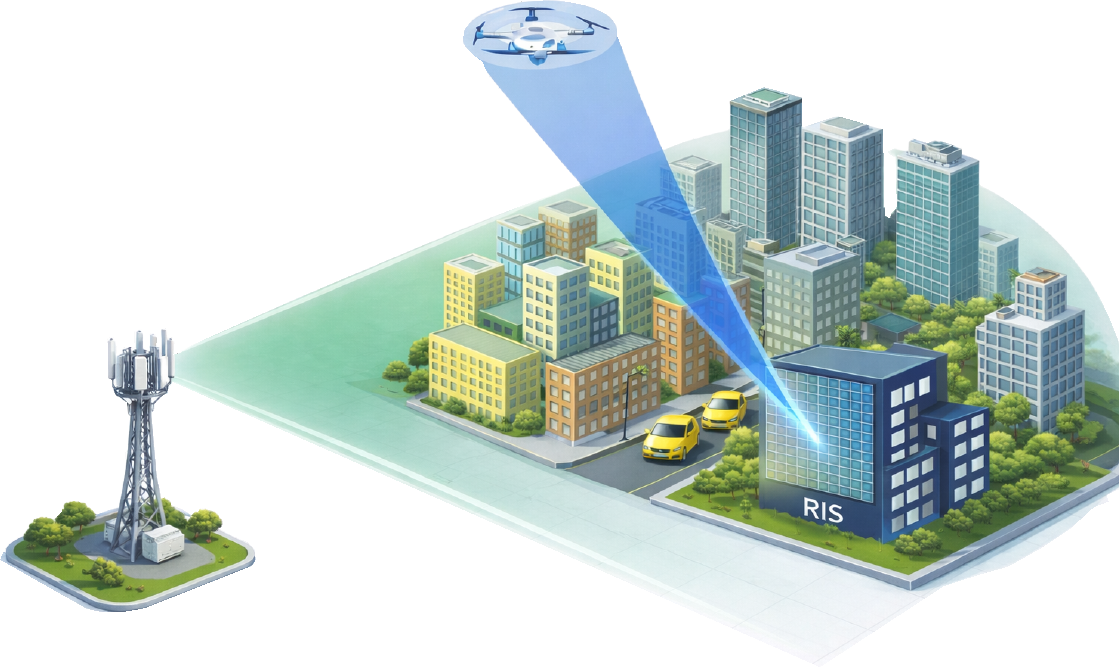}
    \caption{RIS-assisted UAV-based LAWN scenario where the UAV is located outside the BS serving area.}
    \label{fig:System_Model}
\end{figure}
\begin{figure}[t]
    \centering
    \includegraphics[width=\linewidth]{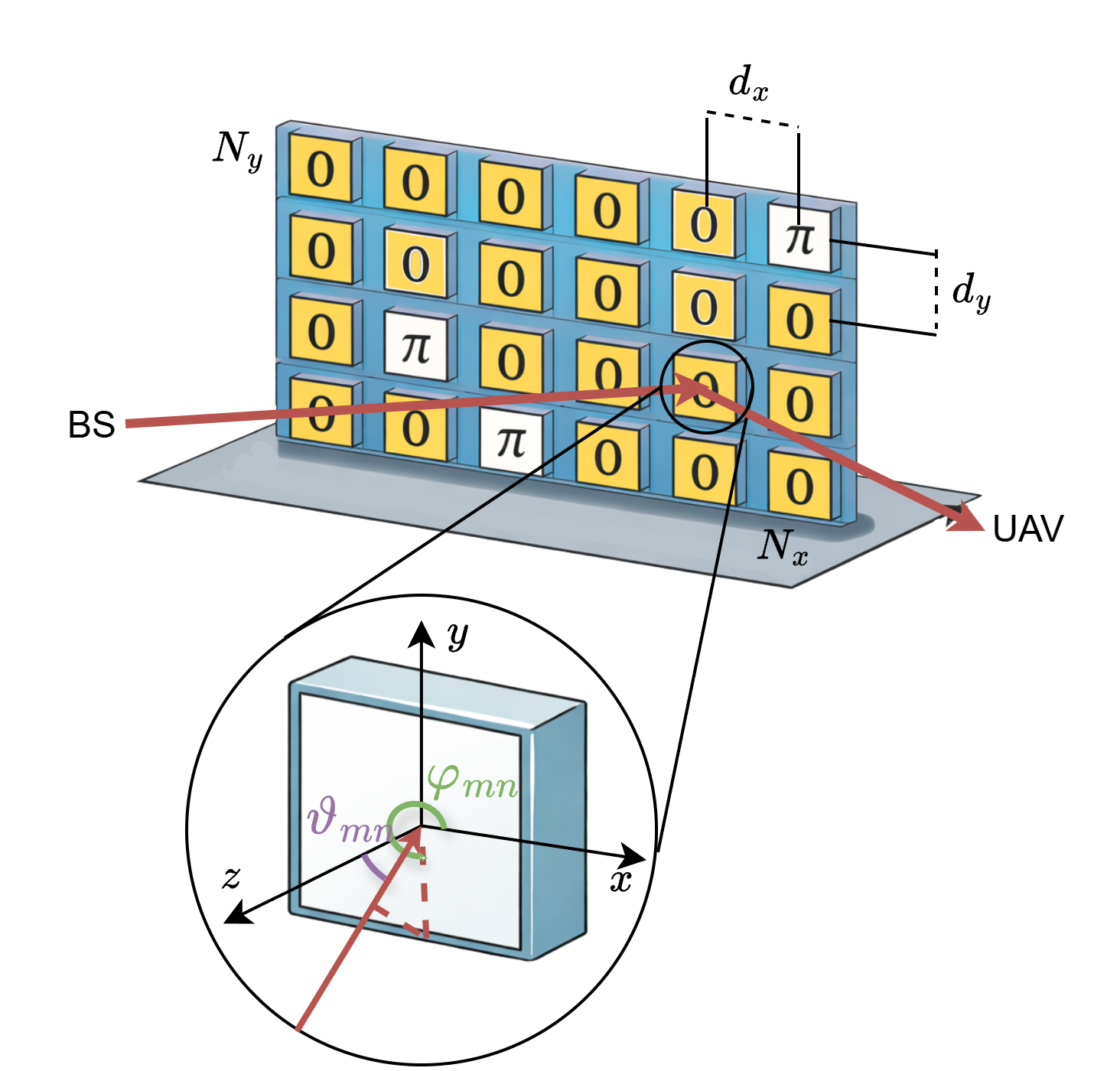}
    \caption{Geometry of the considered RIS model together with the BS-RIS-UAV reflection path and the element-level coordinate frame for the azimuth and elevation angles.}
    \label{fig:RIS_Close}
\end{figure}

An RIS-assisted LAWN is illustrated in Fig.~\ref{fig:System_Model}, in which there is no direct line-of-sight (LOS) link between the base station (BS) and the UAV since the BS's service area is typically configured to serve the ground users. Therefore, wireless communication is established solely through the virtual LOS formed by an RIS. In this scenario, the RIS is modeled as a uniform planar array consisting of $N_x$ elements along the $x$-axis and $N_y$ elements along the $y$-axis, as depicted in Fig.~\ref{fig:RIS_Close}, resulting in a total of $N = N_x N_y$ reflecting elements. All RIS elements lie on the $xy$-plane and face the positive $z$-direction. The complex channel coefficient from the BS and the UAV to the RIS are defined as $\mathbf{h} = [h_{11}, h_{12}, ..., h_{mn}, ..., h_{MN}] \in \mathbb{C}^{N \times 1}$ and $\mathbf{g} = [g_{11}, g_{12}, ..., g_{mn}, ..., g_{MN}]\in \mathbb{C}^{N \times 1}$, respectively. The channel coefficient between the BS and the $(m,n)$th RIS element, $h_{mn}$, is defined as
\begin{equation}\label{eqn:h_mn}
h_{mn} = L_{mn}^{\mathrm{BS}}
\, e^{j k_0 \left(
m d_x \sin\vartheta^{\mathrm{BS}}_{mn} \cos\varphi^{\mathrm{BS}}_{mn}
+n d_y \sin\vartheta^{\mathrm{BS}}_{mn} \sin\varphi^{\mathrm{BS}}_{mn}
\right)},
\end{equation}
while the channel coefficient between the $(m,n)$th RIS element and the UAV, $g_{mn}$, is expressed as
\begin{equation}\label{eqn:g_mn}
g_{mn}=
L_{mn}^{\mathrm{UAV}}
\, e^{j k_0 \left(
m d_x \sin\vartheta^{\mathrm{UAV}}_{mn} \cos\varphi^{\mathrm{UAV}}_{mn}
+n d_y \sin\vartheta^{\mathrm{UAV}}_{mn} \sin\varphi^{\mathrm{UAV}}_{mn}
\right)},
\end{equation}
where $k_0 = 2\pi/\lambda$ denotes the wavenumber, and $d_x$ and $d_y$ are the inter-element spacings along the $x$- and $y$-axes, respectively. The coefficient $L_{mn}^\mathrm{S}$, with $\mathrm{S} \in \{\mathrm{BS}, \mathrm{UAV}\}$, represents the free-space path-loss between node $\mathrm{S}$ and the $(m,n)$th RIS element. The elevation angle with respect to the $(m,n)$th RIS element is denoted by $\vartheta_{mn}^\mathrm{S}$ while the azimuth angle with respect to the $(m,n)$th RIS element is represented by $\varphi_{mn}^\mathrm{S}$~\cite{10200914}. Therefore, the baseband-equivalent received signal at the UAV can be expressed as
\begin{equation}\label{eqn:ReceivedSignal}
y(t)=(\mathbf{h}^\mathrm{T}\mathbf{\Phi}\mathbf{g}) x(t)
+ w(t),
\end{equation}
where $x(t)$ denotes the transmitted signal with $\mathbb{E}\{|x(t)|^2\}=1$, and $w(t) \sim \mathcal{CN}(0,\sigma_w^2)$ represents additive white Gaussian noise. $\mathbf{\Phi}$ denotes the RIS phase shift configuration, defined as $\mathrm{diag}\{\Gamma_{11}e^{j\phi_{11}}, \Gamma_{12}e^{j\phi_{12}}, ..., \Gamma_{mn}e^{j\phi_{mn}}, ..., \Gamma_{MN}e^{j\phi_{MN}} \} \in \mathbb{C}^{N \times N}$. The parameters $\Gamma_{mn}$ and $\phi_{mn}$ represent the reflection amplitude and the additional phase introduced by the $(m,n)$th RIS element, respectively. For simplicity, the reflection amplitude is assumed to be $\Gamma_{mn}=1$ for all elements.

The inter-element spacing is assumed to be half-wavelength, i.e., $d_x=d_y=\lambda/2$, which is commonly used in practical RIS designs. Under this assumption, the effective RIS aperture size can be expressed as
\begin{equation}\label{eqn:D}
    D = \sqrt{
    \left((N_x - 1)\frac{\lambda}{2}\right)^2 +
    \left((N_y - 1)\frac{\lambda}{2}\right)^2 }.
\end{equation}
With the aperture size $D$, the classical electromagnetic region
boundaries are expressed as~\cite{stutzman2012antenna}
\begin{equation}\label{eqn:d_NF-d_FF}
    d_{NF} = 0.62\sqrt{\frac{D^3}{\lambda}}, \qquad
    d_{FF} = \frac{2D^2}{\lambda},
\end{equation}
where $d_{NF}$ and $d_{FF}$ denote the reactive near-field and far-field (Fraunhofer) distances, respectively.

When the BS or the UAV is located in the reactive near-field region ($d < d_{NF}$), where $d$ denotes the distance between the RIS and the BS (or UAV), the electromagnetic wave impinging on the RIS exhibits a spherical wavefront, resulting in a nonlinear phase variation across the RIS surface, as different RIS elements observe distinct propagation distances and angles. In contrast, when the BS or the UAV is located in the far-field region ($d \ge d_{FF}$), the transmitted or reflected electromagnetic wave can be well approximated as a plane wave at the RIS or the UAV, respectively, leading to a linear phase progression along both the $x$ and $y$ axes of the RIS. As a result of this far-field property, the elevation and azimuth angles of the BS and the UAV with respect to the RIS elements can be assumed to be identical across all RIS elements, as follows
\begin{align}
    \vartheta^\mathrm{S} &= \vartheta_{mn}^\mathrm{S}, \quad \forall\, m,n,\;
    \mathrm{S} \in \{\mathrm{UAV},\mathrm{BS}\},\\
    \varphi^\mathrm{S} &= \varphi_{mn}^\mathrm{S}, \quad \forall\, m,n,\;
    \mathrm{S} \in \{\mathrm{UAV},\mathrm{BS}\}.
\end{align}
Accordingly, the phase shift at the UAV arising from both propagation and the phase response of the $(m,n)$th RIS element can be expressed as
\begin{align}\label{eq:Psi}
\Psi_{mn} &=
e^{j k_0 \left(
m d_x \sin\vartheta^{\mathrm{BS}} \cos\varphi^{\mathrm{BS}}
+ n d_y \sin\vartheta^{\mathrm{BS}} \sin\varphi^{\mathrm{BS}}
\right)} \nonumber\\
&\times e^{j\phi_{mn}} \nonumber\\
&\times e^{j k_0 \left(
m d_x \sin\vartheta^{\mathrm{UAV}} \cos\varphi^{\mathrm{UAV}}
+ n d_y \sin\vartheta^{\mathrm{UAV}} \sin\varphi^{\mathrm{UAV}}
\right)} \nonumber\\
&=
e^{j\left(
m k_0 d_x \alpha_x + n k_0 d_y \alpha_y + \phi_{mn}
\right)},
\end{align}
where
\begin{equation}\label{eqn:alphax}
\alpha_x =
\sin\vartheta^{\mathrm{BS}} \cos\varphi^{\mathrm{BS}}
+
\sin\vartheta^{\mathrm{UAV}} \cos\varphi^{\mathrm{UAV}},
\end{equation}
\begin{equation}\label{eqn:alphay}
\alpha_y =
\sin\vartheta^{\mathrm{BS}} \sin\varphi^{\mathrm{BS}}
+
\sin\vartheta^{\mathrm{UAV}} \sin\varphi^{\mathrm{UAV}}.
\end{equation}
The parameters $\alpha_x$ and $\alpha_y$ represent the combined direction components of the BS and UAV links projected onto the RIS surface along the $x$ and $y$ axes, respectively, and determine the phase-gradient across the RIS. Therefore, the phase differences between adjacent RIS elements are constant along both the x-axis and the y-axis, respectively, defined as
\begin{align}
    \Delta_x =k_0 d_x \alpha_x,\label{eqn:Deltax}\\
    \Delta_y =k_0 d_y \alpha_y.\label{eqn:Deltay}
\end{align}
Since the distances from the BS (or UAV) to all RIS elements are nearly equal, resulting in approximately identical path-loss across the surface 
\begin{equation}\label{eq:L}
L^\mathrm{S} \approx L_{mn}^\mathrm{S}, \quad \forall\, m,n, \;
\mathrm{S} \in \{\mathrm{UAV},\mathrm{BS}\}
\end{equation}
where $L_{mn}^\mathrm{S}$ denotes the path-loss between the $(m,n)$th RIS element and the BS (or UAV). Accordingly, the expected SNR of the received signal in~\eqref{eqn:ReceivedSignal} can be written as
\begin{equation}
\mathrm{SNR}
=
\frac{
\left| L^{\mathrm{BS}} L^{\mathrm{UAV}} \right|^2
}{
\sigma_w^2
}
\left|
\sum_{m=0}^{N_x-1} \sum_{n=0}^{N_y-1}
\Psi_{mn}
\right|^2.
\end{equation} 
The received SNR is maximized by selecting the RIS phase shifts so as to maximize the coherent sum of the reflected signals in~\eqref{eq:Psi}. This leads to the following continuous-phase optimization problem:
\begin{equation}\label{eqn:OptimizationProblem}
\begin{aligned}
\max_{\{\phi_{mn}\}} \quad & \left| \sum_{m=0}^{N_x-1} \sum_{n=0}^{N_y-1} e^{j \left( m \Delta_x + n \Delta_y + \phi_{mn} \right)} \right| \\
\text{s.t.} \quad & 0 \leq \phi_{mn} < 2\pi, \quad \forall m, n.
\end{aligned}
\end{equation}
The closed-form solution that achieves coherent phase alignment at the receiver is
\begin{equation}\label{eqn:phi_mn}
\phi_{mn}
=
-  \left( m  \Delta_x + n \Delta_y \right).
\end{equation}
Note that $\Delta_x$ and $\Delta_y$ can be calculated only if the exact 3D positions of the BS and the UAV are known with respect to the RIS. In this study, the UAV position is assumed to be unknown. Under this assumption, the proposed stripe-based algorithm in Section \ref{subsec:Stripe} calculates the RIS phase shifts by efficiently searching different $\Delta_x$ and $\Delta_y$ values, where the received power at the UAV is measured for each pair using \eqref{eqn:phi_mn}.

\section{Stripe-Based RIS Configuration for a Stationary UAV: Optimization and Error Analysis}
This section introduces a stripe-based RIS configuration method for a UAV that hovers or moves negligibly during the RIS optimization interval, allowing its position to be treated as fixed. The proposed formulation is directly derived from the far-field propagation model and admits a low-dimensional parametric representation of the RIS phase profile, which significantly reduces the search space. The method is applicable to both continuous-phase and quantized-phase RIS architectures. In the following, the optimization framework is first presented, and then how phase estimation errors distort the stripe structure and affect the resulting performance is analyzed.

\subsection{Stripe-Based RIS Optimization}\label{subsec:Stripe}
In the stationary far-field UAV scenario, the RIS-assisted link is dominated by a single geometric channel relation between the transmitter, the RIS, and the UAV. Under this condition, the phase profile that maximizes coherent combining across the RIS follows a linear structure over the array surface. This structure enables a reduced-complexity configuration strategy that avoids element-wise phase optimization. From~\eqref{eqn:phi_mn}, the optimal RIS phase varies linearly with the element indices. Hence, the phase map over the surface is fully described by two phase increments along the axes, denoted by $(\Delta_x,\Delta_y)$. The RIS configuration problem is therefore parameterized by this parameter pair instead of element-wise phase variables.

The proposed method searches for the optimal $(\Delta_x,\Delta_y)$ pair rather than directly optimizing individual RIS elements. For any candidate pair, the corresponding phase profile is generated using~\eqref{eqn:phi_mn} and applied to the RIS, and the received signal power $P$ at the UAV  is measured as the objective value. Under continuous-phase control and perfect estimation conditions, this linear phase model represents the optimal far-field solution. However, as the optimal $(\Delta_x,\Delta_y)$ pair must be identified through a search procedure, the resulting solution serves as a near-optimal approximation of the ideal theoretical maximum.

Since $(\Delta_x,\Delta_y)$ are continuous, exhaustive evaluation is not feasible. A two-stage hierarchical search is therefore proposed. In the coarse stage, each dimension is uniformly discretized into $N_c$ points, producing a grid spacing of $\phi_c=\frac{2\pi}{N_c}$. The best coarse point defines a local search window of one grid interval per dimension. A fine grid with $N_f$ levels per dimension is then evaluated inside this window with grid spacing of  $\phi_f=\frac{\phi_c}{N_f}$. The total search cost is $\mathcal{O}(N_c^2+N_f^2)$ power measurements, and the procedure is summarized in Algorithm~\ref{alg:Algorithm1}.

\begin{figure}[t]
    \centering
    \includegraphics[width=\linewidth]{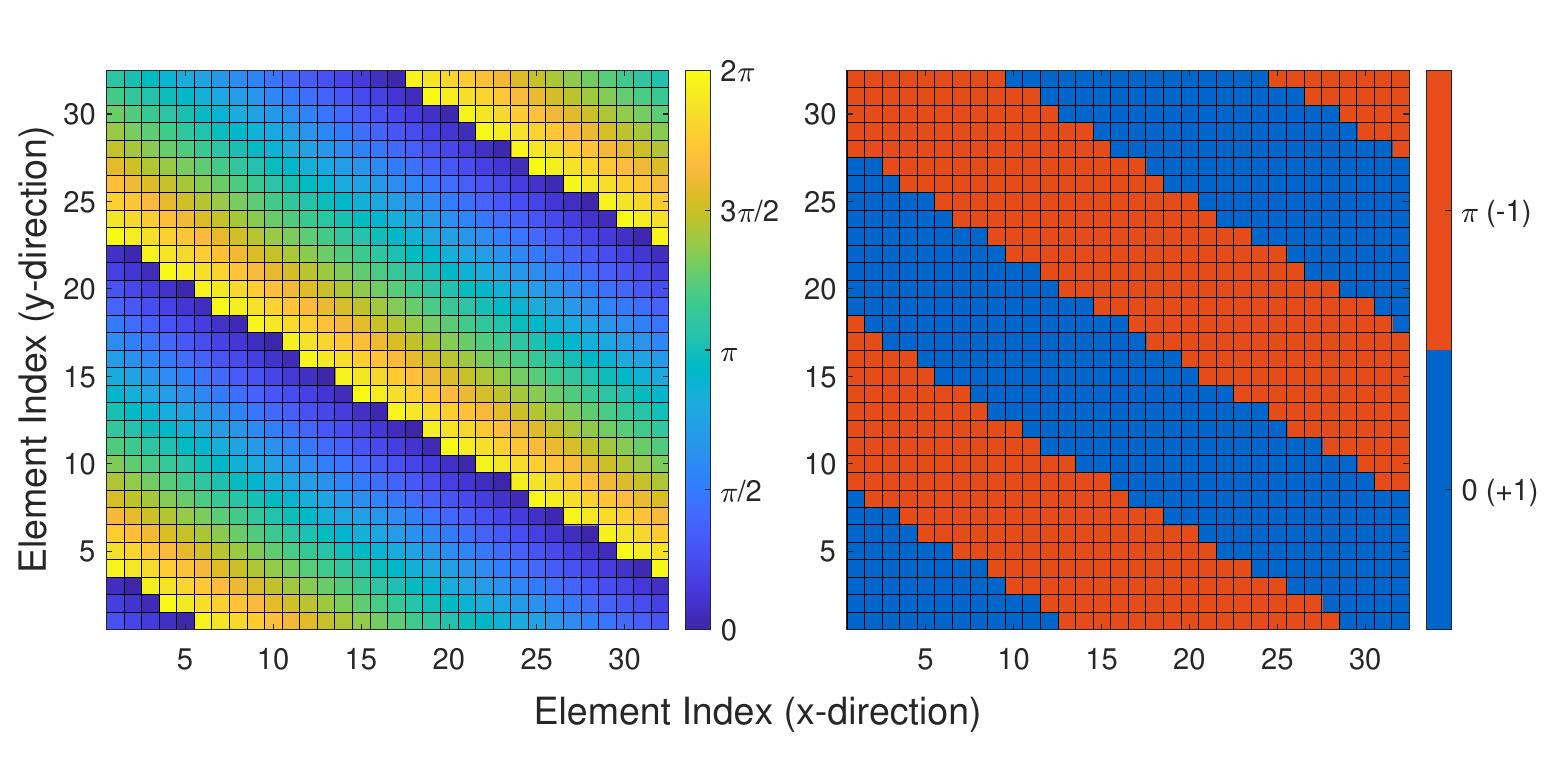}
    \caption{Phase configuration of a $32\times32$ RIS for continuous phases and 1-bit quantized phases, respectively.}
    \label{fig:RIS}
\end{figure}

The same search strategy is applied to the quantized RIS model, and in this work, the system performance is evaluated using a widely adopted and hardware-practical 1-bit RIS whose phase shifts are restricted to the set $\{0,\pi\}$. The 1-bit quantization of $\phi_{mn}$ and $\Psi_{mn}$ are performed as
\begin{equation}\label{eqn:phi^q_mn}
\phi^{(q)}_{mn} =
\begin{cases}
\pi, & \text{if } \mathrm{mod}(\phi_{mn},2\pi) \in [\frac{\pi}{2},\frac{3\pi}{2}),\\
0, &otherwise.
\end{cases}
\end{equation}
\begin{equation}\label{eqn:Psi_mn}
\Psi_{mn}^{(q)} = e^{j\left( m \Delta_x +
n \Delta_y + \phi_{mn}^{(q)}
\right)},
\end{equation}
respectively. Consequently, the linear phase profile is mapped to binary states, which produces stripe-shaped configurations due to phase wrapping, as illustrated in Fig.~\ref{fig:RIS}, considering $(\vartheta^{\mathrm{BS}},\varphi^{\mathrm{BS}})=(60^\circ,25^\circ)$, $(\vartheta^{\mathrm{UAV}},\varphi^{\mathrm{UAV}})=(50^\circ,200^\circ)$, and $(\Delta_x, \Delta_y) \approx (11.71^\circ,18.72^\circ)$. It is worth noting that restricting the search to these stripe patterns avoids exhaustive testing of all $2^{N}$ binary configurations. Instead, the stripe-based optimization needs to find two phase differences, yielding a near-optimum RIS configuration.

\begin{algorithm}[t]
\caption{Hierarchical Stripe-Based RIS Optimization}
\label{alg:Algorithm1}
\begin{algorithmic}[1]
\REQUIRE Coarse resolution $N_c$, fine resolution $N_f$
\ENSURE Optimal phase-gradient pair $(\Delta_x^\star, \Delta_y^\star)$, phase profiles $\phi_{mn}$ and $\phi_{mn}^{(q)}$
\STATE \textbf{Stage 1: Coarse Search}
\STATE Initialize $P_{\max} \leftarrow 0$, $\phi_{c} \leftarrow \frac{2\pi}{N_c}$
\FOR{$i = 0$ \TO $N_c-1$}
    \FOR{$j = 0$ \TO $N_c-1$}
        \STATE $\Delta_x \leftarrow i \phi_c$, \quad $\Delta_y \leftarrow j \phi_c$
        \STATE Apply phase profile using \eqref{eqn:phi_mn} and \eqref{eqn:phi^q_mn} 
        \STATE Measure the received power at the UAV $P_{\mathrm{UAV}}$
        \IF{$P_{\mathrm{UAV}} > P_{\max}$}
            \STATE $P_{\max} \leftarrow P_{\mathrm{UAV}}, \quad (\Delta_x^\star, \Delta_y^\star) \leftarrow (\Delta_x, \Delta_y)$
        \ENDIF
    \ENDFOR
\ENDFOR
\STATE \textbf{Stage 2: Fine search }
\STATE $\phi_{f} \leftarrow \frac{\phi_{c}}{N_f}$ 
\FOR{$i = 0$ \TO $N_f-1$}
    \FOR{$j = 0$ \TO $N_f-1$}
        \STATE $\Delta_x \leftarrow \Delta_x^\star + \left(i - \frac{N_f-1}{2}\right)\phi_f$
        \STATE $\Delta_y \leftarrow \Delta_y^\star + \left(j - \frac{N_f-1}{2}\right)\phi_f$
        \STATE Apply phase profile using \eqref{eqn:phi_mn} and \eqref{eqn:phi^q_mn} 
        \STATE Measure the received power at the UAV $P_{\mathrm{UAV}}$
        \IF{$P_{\mathrm{UAV}} > P_{\max}$}
            \STATE $P_{\max} \leftarrow P_{\mathrm{UAV}}, \quad (\Delta_x^\star, \Delta_y^\star) \leftarrow (\Delta_x, \Delta_y)$
        \ENDIF
    \ENDFOR
\ENDFOR
\STATE Compute final phase profiles $\phi_{mn}$ and $\phi_{mn}^{(q)}$ using \eqref{eqn:phi_mn} and \eqref{eqn:phi^q_mn}, respectively.
\RETURN $\phi_{mn}$ \, $(\phi_{mn}^{(q)})$ 
\end{algorithmic}
\end{algorithm}

\subsection{Phase Estimation Error Analysis}

The practical performance of RIS optimization algorithms depends jointly on their scalability with the RIS size and their robustness to phase estimation errors. In this subsection, the sensitivity of stripe-based 1-bit RIS configurations to phase imperfections is analyzed by explicitly exploiting the structured phase behavior induced under far-field propagation. To isolate the impact of phase estimation inaccuracies on the system performance, these errors are assumed to be the primary source of degradation, and consequently, a noiseless receiver environment is considered. The resulting expressions enable a direct and physically interpretable comparison between structured and unstructured phase errors, thereby clarifying the robustness properties of the proposed stripe-based algorithm.

\subsubsection{Ideal Array Response as a Baseline} 
As established in Section~\ref{SystemModel}, when both the BS and the UAV are located in the far field of the RIS, the cascaded BS-RIS-UAV channel induces a linear phase profile across the RIS aperture. Under this condition, the baseband-equivalent received signal can be written as
\begin{equation}
y(t) = L^{\mathrm{BS}}L^{\mathrm{UAV}}
\sum_{m=0}^{N_x-1}\sum_{n=0}^{N_y-1}
\Psi_{mn} x(t),
\end{equation}
where the ideal continuous RIS phase $\phi_{mn}$ is chosen according to \eqref{eqn:phi_mn} to compensate the deterministic spatial phase terms. With this choice, all RIS elements add coherently at the receiver, yielding
\begin{align}
y_{\mathrm{ideal}}(t) &= L^{\mathrm{BS}}L^{\mathrm{UAV}} N_x N_y \, x(t),
\end{align}
where, $y_{\mathrm{ideal}}$ denotes the received baseband signal under ideal continuous phase control at the RIS, where all reflected components are perfectly phase-aligned at the UAV. Correspondingly, the received signal power under coherent combining at the UAV is defined as
\begin{align}
P_{\mathrm{ideal}} 
&\triangleq \nonumber
\mathbb{E}\!\left[\,|y_{\mathrm{ideal}}(t)|^2\,\right]\\
&=
\left(L^{\mathrm{BS}}L^{\mathrm{UAV}} N\right)^2.
\end{align}

\subsubsection{SNR Sensitivity Analysis Under Phase Difference Estimation Errors}
Stripe-based RIS configurations are fully defined by the parameters $(\Delta_x,\Delta_y)$. Consequently, the dominant source of mismatch originates from errors in estimating these gradients rather than independent element-wise phase perturbations. This effect is modeled by assuming that the phase-gradients are affected by independent estimation errors $e_x$ and $e_y$, such that the gradients become $ \Delta_x + e_x$ and $ \Delta_y + e_y$. Assuming ideal continuous-phase control, since this mismatch manifests itself as a residual linear phase slope across the RIS aperture, the resulting received signal becomes
\begin{align}
\hat{y}_{\mathrm{struct}}(t)
&=
L^{\mathrm{BS}} L^{\mathrm{UAV}}
\sum_{m=0}^{N_x-1}\sum_{n=0}^{N_y-1}
e^{-j(m e_x + n e_y)} x(t) \nonumber\\
&=
L^{\mathrm{BS}} L^{\mathrm{UAV}}
\left(\sum_{m=0}^{N_x-1} e^{-j m e_x}\right)
\left(\sum_{n=0}^{N_y-1} e^{-j n e_y}\right) x(t),
\end{align}
where each summation corresponds to a finite geometric series. The resulting received power can therefore be expressed as
\begin{equation}
P_{\mathrm{struct}}
=
\left(L^{\mathrm{BS}} L^{\mathrm{UAV}} N\right)^2
G_x^2(e_x)\,G_y^2(e_y),
\end{equation}
with
\begin{align}
G_x(e_x)=
\frac{1}{N_x}
\frac{\sin(N_x e_x/2)}{\sin(e_x/2)},
\qquad
G_y(e_y)=
\frac{1}{N_y}
\frac{\sin(N_y e_y/2)}{\sin(e_y/2)}.
\end{align}
In practical RIS implementations, the applied phase shifts are restricted to a 1-bit alphabet, as defined in \eqref{eqn:phi^q_mn}. The impact of quantization on the average coherent gain reduction can be expressed as~\cite{8930608}
\begin{equation}
\mathrm{Q_{loss}}
=
10\log_{10}\!\left(\frac{4}{\pi^2}\right).
\end{equation}
Accordingly, the received power in the presence of structured phase-gradient errors and 1-bit phase quantization can be approximated as
\begin{equation}
P_{\mathrm{struct}}^{(q)}
=
\frac{4}{\pi^2}
\left(L^{\mathrm{BS}} L^{\mathrm{UAV}} N\right)^2
G_x^2(e_x)\,G_y^2(e_y).
\end{equation}
The corresponding power loss with respect to the ideal continuous-phase RIS configuration is then given by
\begin{align}
\mathcal{L}_{\mathrm{struct}}^{\mathrm{(dB)}}
=&
10\log_{10}\!\left(
\frac{P_{\mathrm{ideal}}}{P_{\mathrm{struct}}^{(q)}}
\right)\\\nonumber
=&
-\mathrm{Q_{loss}}
-
10\log_{10}\!\left(G_x^2(e_x)G_y^2(e_y)\right).
\end{align}

\subsubsection{Random Phase Errors}
As a benchmark, a scenario is considered in which each RIS element independently applies an incorrect 1-bit phase with probability $p$, thereby eliminating the spatial correlation induced by the far-field geometry. Specifically, let $b_{mn} \in \{0,1\}$ be an independent Bernoulli random variable with $\Pr(b_{mn}=1)=p$. The applied random phase factor for the $(m,n)$th RIS element is given by $e^{j\pi b_{mn}}$. Accordingly, the baseband-equivalent received signal is expressed as
\begin{equation}
    \hat{y}_{\mathrm{rand}}(t) = L^{\mathrm{BS}} L^{\mathrm{UAV}} \sum_{m=0}^{N_x-1} \sum_{n=0}^{N_y-1} \Psi_{mn}^{(q)} e^{j\pi b_{mn}} x(t).
\end{equation}
The average received power $\mathbb{E}[P_{\mathrm{rand}}^{(q)}]$ is decomposed into a coherent part and an incoherent power floor. The statistical properties of the phase factor are defined as:
\begin{align}
    \mathbb{E}[e^{j\pi b_{mn}}] = 1-2p, \qquad
    \mathrm{Var}(e^{j\pi b_{mn}}) = 4p(1-p).
\end{align}
Using these properties, the expected power is derived as
\begin{align}
\mathbb{E}[P_{\mathrm{rand}}^{(q)}] &= (L^{\mathrm{BS}} L^{\mathrm{UAV}})^2 \nonumber\\
&\times\left( \left| (1-2p) \sum_{m,n} \Psi_{mn}^{(q)} \right|^2 + \sum_{m,n} 4p(1-p) \right) \nonumber \\
&= (L^{\mathrm{BS}} L^{\mathrm{UAV}} N)^2 \left[ \frac{1}{N} + \left( \frac{4}{\pi^2} - \frac{1}{N} \right) (1-2p)^2 \right].
\end{align}
The corresponding power loss in dB relative to the ideal continuous-phase configuration is
\begin{equation}
\mathcal{L}_{\mathrm{rand}}^{\mathrm{(dB)}} = -10\log_{10} \left( \frac{1}{N} + \left( \frac{4}{\pi^2} - \frac{1}{N} \right) (1-2p)^2 \right).
\end{equation}

To enable a meaningful comparison between structured and random errors, a mapping between the deterministic phase excursion $\Theta_{\mathrm{exc}}$ and the probability $p$ is established. The total phase excursion is defined as $\Theta_{\mathrm{exc}} = e_x(N_x-1) + e_y(N_y-1)$ and is associated with an equivalent inversion probability:
\begin{equation}
\label{eq:p_e_mapping}
p = \frac{\Theta_{\mathrm{exc}}}{2\pi}.
\end{equation}
By substituting \eqref{eq:p_e_mapping} into the loss expression, the average power loss under random phase errors can be expressed compactly in terms of the total phase excursion as:
\begin{equation}
\mathcal{L}_{\mathrm{rand}}^{\mathrm{(dB)}} = -10\log_{10} \left[ \frac{1}{N} + \left( 1 - \frac{1}{N} \right) \frac{4}{\pi^2} \left( 1 - \frac{\Theta_{\mathrm{exc}}}{\pi} \right)^2 \right].
\end{equation}
This expression explicitly captures the combined impact of uncorrelated phase flips, 1-bit phase quantization, and RIS aperture size, allowing for a direct comparison with $\mathcal{L}_{\mathrm{struct}}^{\mathrm{(dB)}}$. 

\section{ Stripe-Based RIS Configuration For a Moving UAV: Tracking and Angular Localization}
Unlike the stationary case, where the RIS configuration is computed once, a mobile UAV necessitates continuous adaptation of the phase-gradients to maintain the beamforming gain. However, performing a search over the entire configuration space at every time step is computationally prohibitive for real-time applications due to processing latency and finite RIS controller update rates.

To address this challenge, temporal correlations in the UAV's trajectory are leveraged. Since the UAV's position changes continuously, the optimal RIS parameters at any given time are constrained by the previous state and the physical limits of motion. In this section, the dynamics of the phase gradients under mobility are first characterized to establish a bound on the search space, which then forms the basis for the proposed tracking strategies.

\subsection{Tracking Algorithms}
To design efficient tracking updates, it is first necessary to model how the UAV's physical motion translates into shifts in the optimal $(\Delta_x ,\Delta_y)$ phase-gradients. Consider a UAV moving with velocity $\mathbf{v}(t)$ and position $\mathbf{r}(t)$ with respect to the RIS coordinate system at time instance $t$, defined as:
\begin{equation}
\mathbf{v}(t)=
\begin{bmatrix}
v_x(t)\\ v_y(t)\\ v_z(t)
\end{bmatrix},
\qquad
\mathbf{r}(t)=
\begin{bmatrix}
x(t)\\ y(t)\\ z(t)
\end{bmatrix}.
\end{equation}
The UAV's position at any time $\mathbf{r}(t_0+t)$ is governed by its initial state and the integral of its velocity through
\begin{equation}
\mathbf{r}(t_0+t)=\mathbf{r}(t_0)+\int_{t_0}^{t_0+t}\mathbf{v}(\tau)\,d\tau.
\end{equation}

In the proposed stripe-based configuration, the required combined direction components $\alpha_x(t)$ and $\alpha_y(t)$ are the sum of a static component (from the fixed BS) and a time-varying component (from the mobile UAV):
\begin{align}
\alpha_x(t) &= \alpha_x^{\mathrm{BS}} + \alpha_x^{\mathrm{UAV}}(t), \\
\alpha_y(t) &= \alpha_y^{\mathrm{BS}} + \alpha_y^{\mathrm{UAV}}(t).
\end{align}
The BS-induced components, $\alpha_{x}^{\mathrm{BS}}$ and $ \alpha_{y}^{\mathrm{BS}}$, depend solely on the fixed direction of arrival and remain constant. Thus, the temporal drift in the RIS configuration is driven entirely by the UAV-induced terms, which are functions of the unit direction vector $\mathbf{u}(t) = \mathbf{r}(t)/R(t)$, where $R(t)=\|\mathbf{r}(t)\|$:
\begin{equation}
\alpha_x^{\mathrm{UAV}}(t)=\frac{x(t)}{R(t)}=u_x(t),\qquad
\alpha_y^{\mathrm{UAV}}(t)=\frac{y(t)}{R(t)}=u_y(t).
\end{equation}
To quantify the rate of change in these parameters, the UAV-induced terms are differentiated with respect to time, yielding:
\begin{align}
\dot{\alpha}_x(t) &= \frac{1}{R(t)} \left( v_x(t) - u_x(t)\big(\mathbf{u}(t)\cdot\mathbf{v}(t)\big) \right), \\
\dot{\alpha}_y(t) &= \frac{1}{R(t)} \left( v_y(t) - u_y(t)\big(\mathbf{u}(t)\cdot\mathbf{v}(t)\big) \right).
\end{align}
These expressions reveal a key physical insight: only velocity components orthogonal to the RIS-UAV line induce changes in the combined direction parameters, whereas radial motion only scales the propagation distance. 

For algorithmic implementation, instead of instantaneous derivatives, the accumulated variation over a tracking interval $\delta_t$ is considered. Assuming that the UAV speed is bounded as $\|\mathbf{v}(t)\|\le v_{\max}$, the drift in the combined direction parameters is upper bounded by
\begin{equation}
\max_{i \in \{x,y\}}
\left|
\alpha_i(t_0+\delta_t) - \alpha_i(t_0)
\right|
\le
\frac{v_{\max}}{R_{\min}}\,\delta_t,
\end{equation}
where $R_{\min}$ denotes the minimum RIS-UAV distance. Since the phase-gradients are defined as $\Delta_i = k_0 d_i \alpha_i$, this directly yields the phase-gradient drift bound
\begin{equation}
\left| \Delta_i(t_0+\delta_t) - \Delta_i(t_0)
\right| \le k_0 d_i \frac{v_{\max}}{R_{\min}}\,\delta_t,
\quad i\in\{x,y\}.
\end{equation}
Therefore, instead of searching over the full $(0,2\pi]$ domain, the search space of $\Delta_x$ and $\Delta_y$ can be restricted to a local interval around the previous estimate as
\begin{align}
\Delta_i(t_0+\delta_t) \in \left[ \Delta_i(t_0) \pm
k_0 d_i \frac{v_{\max}}{R_{\min}}\,\delta_t \right],
\quad i\in\{x,y\}.
\end{align}
Based on this dynamic characterization, three complementary tracking strategies are considered.

\subsubsection{Simple Tracking}
The first tracking strategy is based on repeatedly executing the stripe-based optimization procedure in Algorithm~\ref{alg:Algorithm1} using only the fine-resolution mode. While the original algorithm employs a coarse-to-fine strategy, the coarse stage becomes unnecessary in a tracking context once an initial estimate of $(\Delta_x,\Delta_y)$ is available. Consequently, the optimization is continuously operated at fine resolution to refine the estimate around the previously identified optimum. In each tracking cycle, $N_f^2$ candidate stripe configurations are tested sequentially. Each configuration is applied for a measurement duration $\delta_m$, yielding the total update interval
\begin{equation}
\delta_u = N_f^2 \delta_m.
\end{equation}
Due to UAV mobility, the true phase-gradient parameters vary over this interval. Using the finite-time drift bound derived earlier, the worst-case variation satisfies
\begin{equation}
|\Delta_i(t_k+\delta_u) - \Delta_i(t_k)|
\le
k_0 d_i \frac{v_{\max}}{R_{\min}}\,\delta_u,
\qquad i\in\{x,y\}.
\end{equation}
To ensure that the true optimum remains inside the explored region throughout one update interval, the search half-width must satisfy
\begin{equation}
W \ge
\max_{i \in \{x,y\}}
\left(
k_0 d_i \frac{v_{\max}}{R_{\min}}\,\delta_u
\right).
\end{equation}
In practice, $W$ is selected according to this bound (optionally with a safety margin) and kept fixed during tracking. Let $(\Delta_x(t_k), \Delta_y(t_k))$ denote the estimated optimal phase-gradient pair at the start of the $k$-th cycle at time $t_k$, and define the index set
\[
\mathcal{I} =
\left\{-\frac{N_f-1}{2}, \dots, \frac{N_f-1}{2}\right\}.
\]
A local discrete search grid satisfying this window constraint is constructed as
\begin{equation}
\mathcal{G}_k =
\Big\{
\big(
\Delta_x(t_k) + r\varrho,\;
\Delta_y(t_k) + \ell\varrho
\big)
\Big\}_{r,\ell \in \mathcal{I}},
\end{equation}
where the grid spacing is defined as
\begin{equation}
\varrho= \frac{2W}{N_f-1}.
\end{equation}
After evaluating all $N_f^2$ candidates in $\mathcal{G}_k$, the phase-gradient pair that yields the largest measured received signal power among the tested configurations is selected, and its corresponding $(\Delta_x,\Delta_y)$ values are taken as
$(\Delta_x(t_{k+1}),\Delta_y(t_{k+1}))$ for the next cycle.

\subsubsection{Kalman Aided Tracking}
The second strategy improves upon the simple tracking method by explicitly modeling the temporal evolution of the phase-gradient parameters and exploiting correlations across successive update cycles. Rather than treating each grid-search outcome independently, the sequence of estimated parameters is assumed to follow an underlying discrete-time dynamic process. Let the state vector be defined as
\begin{equation}
\mathbf{s}(t_k)=
\begin{bmatrix}
\Delta_x(t_k)\\
\Delta_y(t_k)\\
\dot{\Delta}_x(t_k)\\
\dot{\Delta}_y(t_k)
\end{bmatrix},
\end{equation}
whose evolution over one update interval $\delta_u$ is modeled as
\begin{equation}
\mathbf{s}(t_{k+1})
=
\mathbf{K}\,\mathbf{s}(t_k) + \mathbf{n}(t_k),
\end{equation}
where $\mathbf{K}$ is defined as
\begin{equation}
    \mathbf{K} = \begin{bmatrix}
        1 & 0 & \delta_u & 0\\
        0 & 1 & 0 & \delta_u\\
        0 & 0 & 1 & 0\\
        0 & 0 & 0 & 1\\       
    \end{bmatrix},
\end{equation}
and $\mathbf{n}(t_k)$ is a zero-mean process noise term accounting for unmodeled dynamics such as deviations from constant velocity, small trajectory curvature, and abrupt UAV maneuvers.

Based on this model, a prediction
$(\hat{\Delta}_x^{-},\hat{\Delta}_y^{-})$ is generated prior to the next grid-search cycle. The subsequent grid search is centered around this prediction,
\begin{equation}
\Delta_x \in [\hat{\Delta}_x^{-}-W_p,\hat{\Delta}_x^{-}+W_p],\qquad
\Delta_y \in [\hat{\Delta}_y^{-}-W_p,\hat{\Delta}_y^{-}+W_p],
\end{equation}
where $W_p$ denotes the prediction-based search window. Since the prediction incorporates information from multiple past cycles, the window $W_p$ can typically be chosen such that
\begin{equation}
W_p < W,
\end{equation}
leading to smaller SNR fluctuations during the search. After each cycle, the phase-gradient pair selected by the grid search is treated as a measurement and used to update the state estimate through a Kalman filtering step.

\subsubsection{Extremum Seeking Control Aided Tracking}
The third strategy adopts an ESC framework to track the phase-gradients without explicit grid searches. For notational simplicity, let the received power measured at time index $t$ be expressed as an unknown nonlinear function
\begin{equation}
P(t)\triangleq P(\Delta_x(t),\Delta_y(t)).
\end{equation}
Bounded small perturbations are superimposed onto the phase-gradients,
\begin{align}
\Delta_x(t) &= \bar{\Delta}_x(t) + a_x \sin(\omega_x t),\\
\Delta_y(t) &= \bar{\Delta}_y(t) + a_y \sin(\omega_y t),
\end{align}
where $a_x,a_y$ denote the perturbation amplitudes, $\omega_x,\omega_y$ are distinct perturbation frequencies. The received power is sampled every $\delta_m$ seconds.
To estimate the local gradient of the received power, the power measurements are correlated with the perturbation signals over a finite averaging window length $N_a$,
\begin{align}
\hat{g}_x(t) &=
\frac{2}{a_x}\frac{1}{N_a}
\sum_{i=0}^{N_a-1}
\big(P(t-i\delta_m)-\bar{P}(t)\big)
\sin\!\big(\omega_x (t-i\delta_m)\big),\\
\hat{g}_y(t) &=
\frac{2}{a_y}\frac{1}{N_a}
\sum_{i=0}^{N_a-1}
\big(P(t-i\delta_m)-\bar{P}(t)\big)
\sin\!\big(\omega_y (t-i\delta_m)\big),
\end{align}
where $\bar{P}(t)$ denotes the local average power over the time period $N_a\delta_m$. The nominal phase-gradient parameters are updated according to
\begin{align}
\bar{\Delta}_x(t+\delta_m) &= \bar{\Delta}_x(t) + \kappa\,\hat{g}_x(t),\\
\bar{\Delta}_y(t+\delta_m) &= \bar{\Delta}_y(t) + \kappa\,\hat{g}_y(t),
\end{align}
where $\kappa>0$ is the adaptation gain controlling the trade-off between tracking speed and noise robustness. This approach enables continuous low-latency adaptation without requiring predefined search windows or grid-based exploration.

\subsection{UAV Tracking}
Beyond RIS configuration optimization, the proposed optimization framework also enables localization of the UAV in terms of its angular direction relative to the RIS. This capability follows directly from the physical interpretation of the combined direction parameters $(\alpha_x,\alpha_y)$ estimated during the optimization process.

Recalling the definitions in \eqref{eqn:alphax} and \eqref{eqn:alphay}, the effective combined direction parameters can be decomposed into BS-induced and UAV-induced components. Rearranging these expressions yields
\begin{align}
\sin\vartheta^{\mathrm{UAV}}(t)\cos\varphi^{\mathrm{UAV}}(t)
&=
\alpha_x(t)-\alpha_x^{\mathrm{BS}}, \label{eqn:dx_loc}\\
\sin\vartheta^{\mathrm{UAV}}(t)\sin\varphi^{\mathrm{UAV}}(t) 
&=
\alpha_y(t)-\alpha_y^{\mathrm{BS}}, \label{eqn:dy_loc}
\end{align}
where $\vartheta^{\mathrm{UAV}}(t)$ and $\varphi^{\mathrm{UAV}}(t)$ denote the elevation and azimuth angles of the UAV with respect to the RIS, respectively, and $(\alpha_x^{\mathrm{BS}},\alpha_y^{\mathrm{BS}})$ are constant and known for a fixed BS location.
Taking the ratio of \eqref{eqn:dy_loc} and \eqref{eqn:dx_loc} allows direct estimation of the UAV azimuth angle,
\begin{equation}
\varphi^{\mathrm{UAV}}(t)
=
\arctan\!\left(
\frac{\alpha_y(t)-\alpha_y^{\mathrm{BS}}}
{\alpha_x(t)-\alpha_x^{\mathrm{BS}}}
\right).
\end{equation}
Similarly, squaring and summing \eqref{eqn:dx_loc} and \eqref{eqn:dy_loc}, and using the identity
$\sin^2\vartheta=\sin^2\vartheta(\cos^2\varphi+\sin^2\varphi)$, yields the elevation angle
\begin{equation}
\vartheta^{\mathrm{UAV}}(t)
=
\arcsin\!\left(
\sqrt{
\left(\alpha_x(t)-\alpha_x^{\mathrm{BS}}\right)^2
+
\left(\alpha_y(t)-\alpha_y^{\mathrm{BS}}\right)^2
}
\right).
\end{equation}

These expressions indicate that, once the $(\alpha_x(t),\alpha_y(t))$ parameters are estimated, the UAV direction can be inferred without any additional signaling, training sequences, or explicit localization procedures. Importantly, this angular information is obtained as a byproduct of RIS optimization, rather than through a dedicated localization algorithm.

In practice, however, the optimization process may involve intentional beam exploration or local searches that do not always point toward the true UAV direction. Consequently, not all intermediate angle estimates are equally reliable. To ensure robust localization, only the combined direction parameter estimates corresponding to the maximum received SNR are retained and used for UAV tracking. This selection mechanism naturally filters out spurious estimates and aligns the localization output with the physically dominant propagation direction.

Overall, this result highlights an additional benefit of the proposed algorithm. In addition to enabling low-complexity and robust RIS optimization, it provides implicit UAV angular localization capabilities that are particularly attractive for mobile and vehicular communication scenarios.

\section{Performance Evaluation}
In this section, the proposed algorithm and its performance under different scenarios are assessed through MATLAB simulations. Unless otherwise stated, in the simulation setup, the RIS has a $24\times24$ ($N=576$-elements) operating at $3.6$ GHz. The Monte Carlo simulations of $1000$ random realizations are performed to obtain the average results. The average SNR loss results obtained through analysis and simulation are closely matched, verifying the phase estimation error analysis of the stripe-based method. To validate the underlying assumptions and the practical viability of the proposed algorithm, measurement experiments are conducted with a real RIS prototype in an outdoor campus environment using a stationary software-defined radio. 

\subsection{SNR Results for Varying UAV Distances}
\begin{figure}[t]
    \centering
    \includegraphics[width=\linewidth]{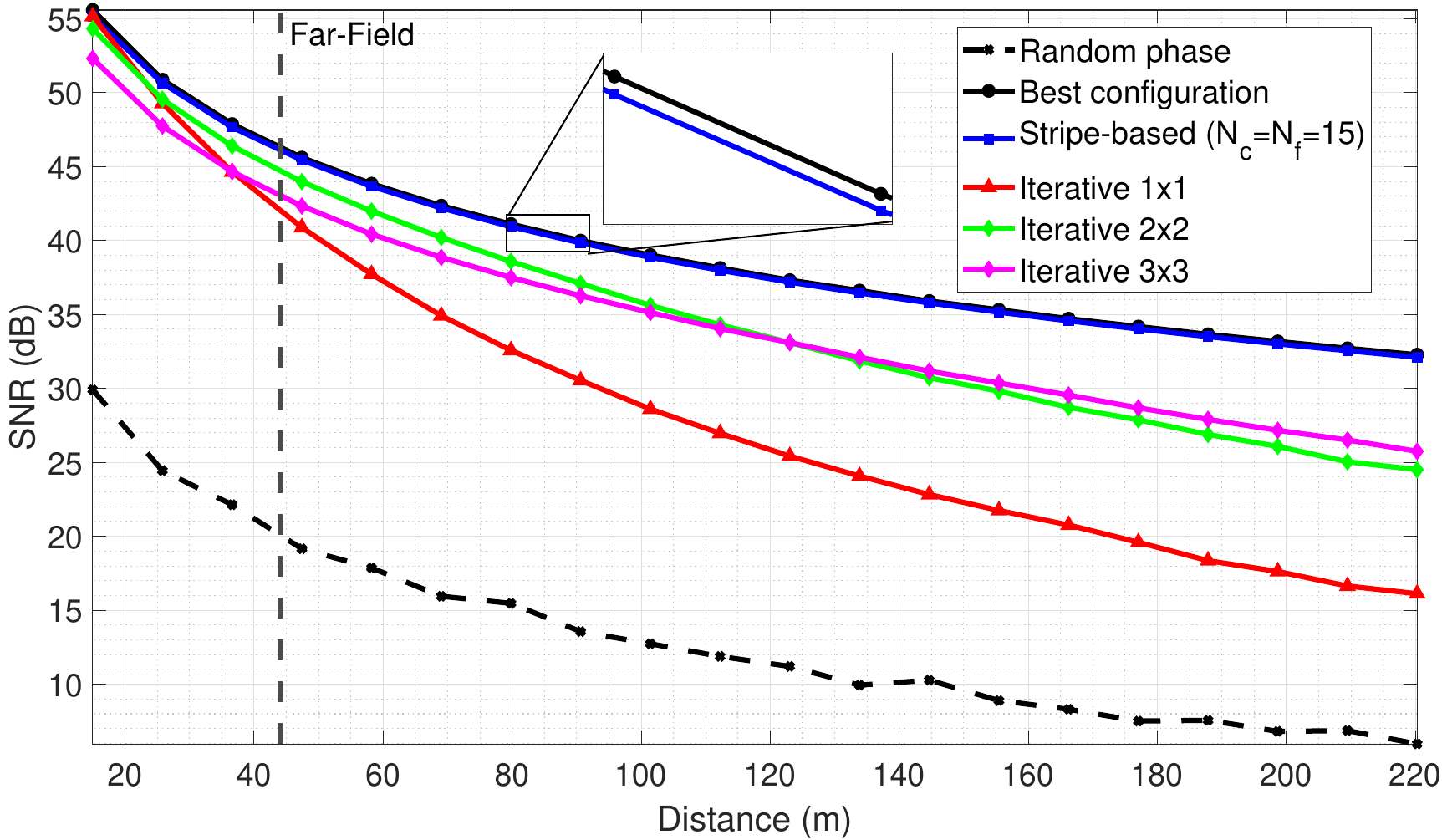}
    \caption{The simulation results of the average received SNR for varying UAV-RIS distances.}
    \label{fig:FarNearField_60dBm}\vspace{-10pt}
\end{figure}

The distance sensitivity of several RIS configuration strategies is investigated in Fig.~\ref{fig:FarNearField_60dBm}. The RIS is centered at the origin, while the BS is positioned $100$ m away from the RIS and transmits with a power of $60$ dBm. The UAV is positioned at radial distances ranging from $d_{FF}/3$ to $5d_{FF}$ from the RIS, where $d_{FF} = 44$ m in the simulation setup, while maintaining fixed azimuth and elevation angles of $20^{\circ}$. The noise power at the UAV is set to $-95$ dBm. For comparison, the random-phase baseline and the practical upper bound are evaluated for $1$-bit RIS configurations.
While the conventional $1\times 1$ iterative method~\cite{10278759} is near-optimal at short ranges, its SNR degrades rapidly as distance increases. This performance drop occurs because, even when the total received SNR is high, the impact of a single element's phase change can be too small to be distinguished from the noise, preventing accurate phase updates. Conversely, larger blocks ($2\times 2$ and $3\times 3$) improve robustness to distance by averaging decisions across multiple elements, providing more reliable updates at the expense of lower spatial resolution. Therefore, while these methods underperform the $1\times 1$ case at short ranges, they remain more stable as distance increases. Among them, the $2\times 2$ scheme is preferable at moderate distances, whereas the $3\times 3$ scheme becomes more effective at longer ranges due to its higher noise tolerance.

In contrast, the proposed stripe-based algorithm ($N_c=N_f=15$) consistently achieves high SNR and closely follows the practical upper bound across all distances. This superiority stems from its hybrid nature: it updates multiple elements simultaneously to ensure that phase adjustments remain detectable above the noise floor, while still allowing the RIS to reach a fine-grained final configuration. Consequently, the proposed method effectively combines the noise robustness of block-based schemes with the high spatial resolution of element-wise optimization.

\begin{figure}[t]
    \centering
    \includegraphics[width=\linewidth]{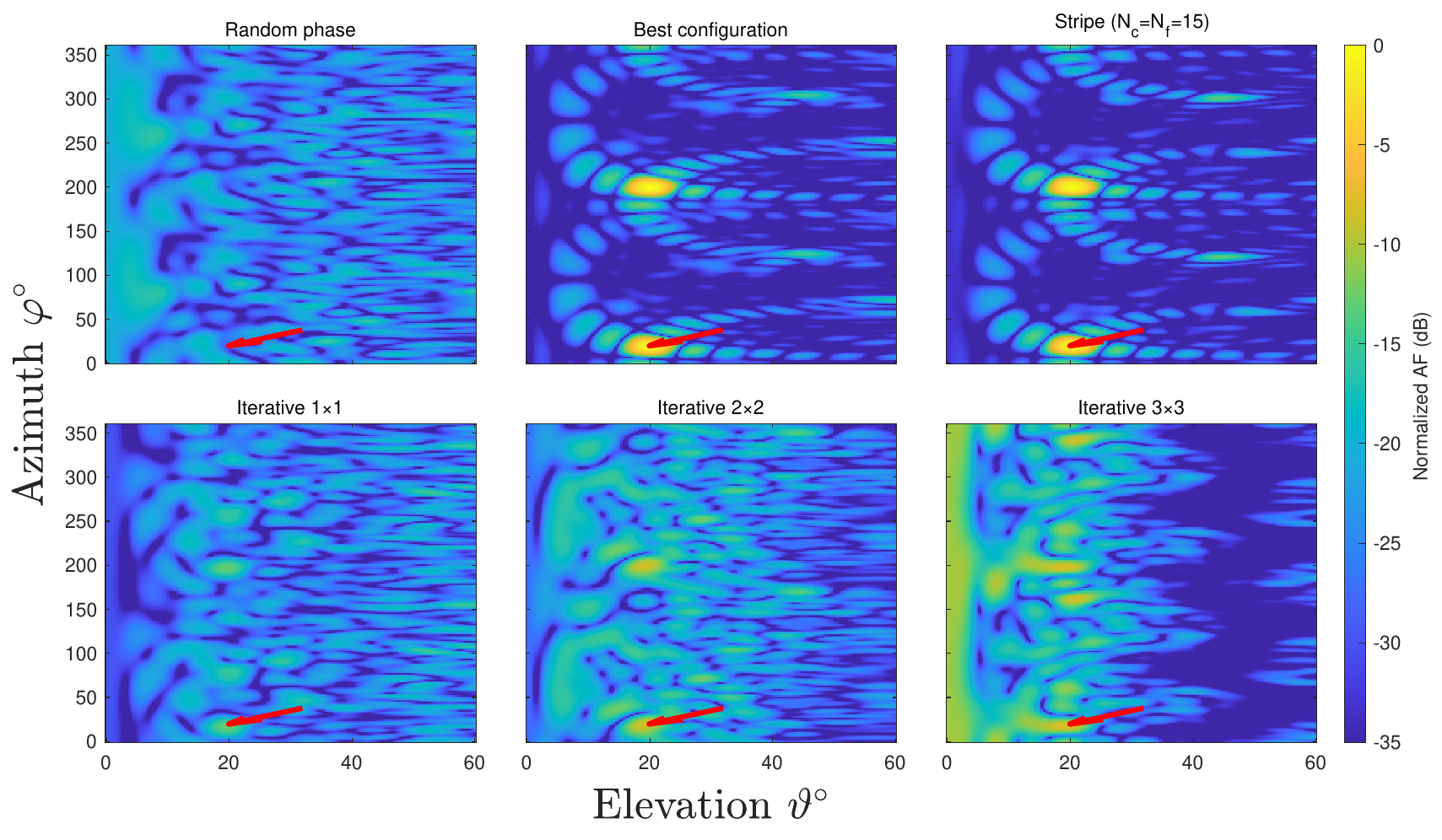}
    \caption{Radiation pattern of the reflected signal from RIS to the UAV.}
    \label{fig:Radiation_Pattern_Normalized}\vspace{-10pt}
\end{figure}

Fig. \ref{fig:Radiation_Pattern_Normalized} compares the normalized radiation patterns for different RIS configuration methods. The results are presented as an example corresponding to a single Monte Carlo realization, where the UAV is located at a distance of $5d_{FF}$. The red arrows indicate the target UAV direction. As observed, the proposed stripe-based algorithm achieves a sharp main lobe that closely matches the best configuration, indicating high beamforming efficiency and directivity. In contrast, the $1\times 1$ iterative method shows a highly dispersed radiation pattern with significant gain loss. This behavior confirms that single-element updates are easily submerged in noise, leading to inaccurate phase focusing. The stripe-based approach overcomes this limitation by updating element groups (stripes), ensuring that the phase contribution of each iteration remains distinguishable from the noise while maintaining a near-optimal focus at the target receiver.

\subsection{Runtime Results}
\begin{figure}[t]
    \centering
    \includegraphics[width=1\linewidth]{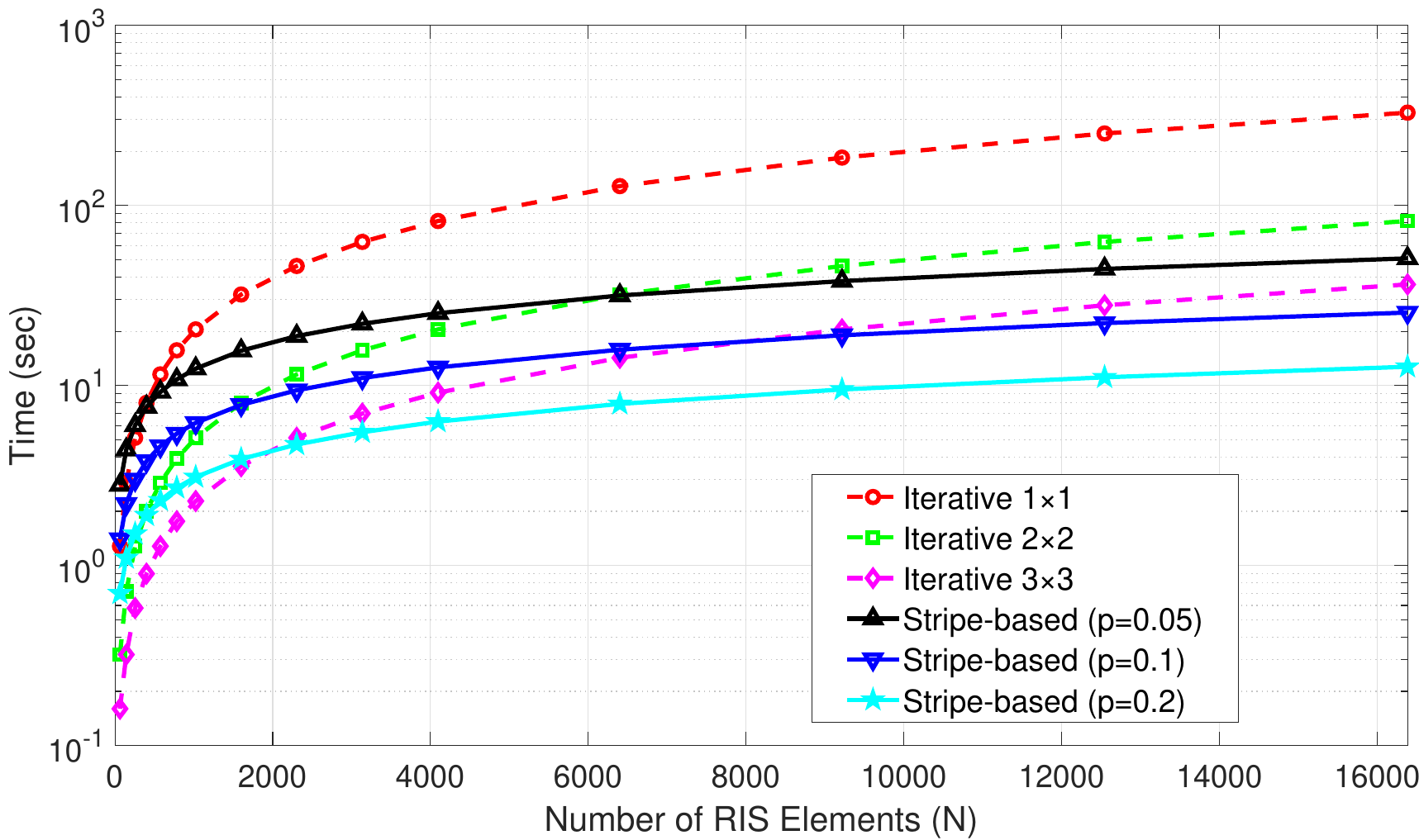}
    \caption{The runtime comparison of the RIS optimization methods. Iterative methods (dashed) scale as $\mathcal{O}(N)$, while proposed stripe-based (solid) achieves $\mathcal{O}(\sqrt{N})$ for constant bit flip rate p.}
    \label{fig:Complexity_Analysis}
\end{figure}

The optimization runtime as a function of RIS sizes for the iterative and stripe-based methods is illustrated in Fig.~\ref{fig:Complexity_Analysis}. A square RIS is considered with $N=N_x N_y$ elements and sweep $N_x$ from $8$ to $128$. The number of configuration evaluations is converted to time by assuming a fixed measurement duration of $10$ ms per evaluation. For the stripe-based algorithm, a constant bit-flip error ratio $p\in\{0.05, 0.1, 0.2\}$ is enforced as the RIS size grows, where $p$ denotes the fraction of RIS elements that end up flipped incorrectly after the optimization. To keep this ratio constant for larger arrays, the resolution parameters $(N_c,N_f)$ are increased with RIS size according to \eqref{eq:p_e_mapping} to reduce the $e_x$ and $e_y$. In contrast, the iterative baselines use fixed element-wise ($1\times1$) and block-wise ($2\times2$, $3\times3$) update rules, which do not scale with RIS size.

The figure indicates distinct scaling behaviors. The iterative methods require a number of evaluations proportional to the total number of RIS elements, i.e., linear in $N$, so their optimization time increases linearly with surface size. Under constant bit-flip error ratio operation, the stripe-based optimization method requires a number of evaluations proportional to $N_x+N_y$ (equivalently proportional to $\sqrt{N}$ for a square RIS), resulting in substantially slower growth with RIS size. While decreasing $p$ increases the required evaluations and thus the optimization time, the stripe-based algorithm still exhibits significantly slower growth with RIS size than the iterative.

\subsection{Effects of Phase Estimation Error}
\begin{figure}[t]
    \centering
    \includegraphics[width=\linewidth]{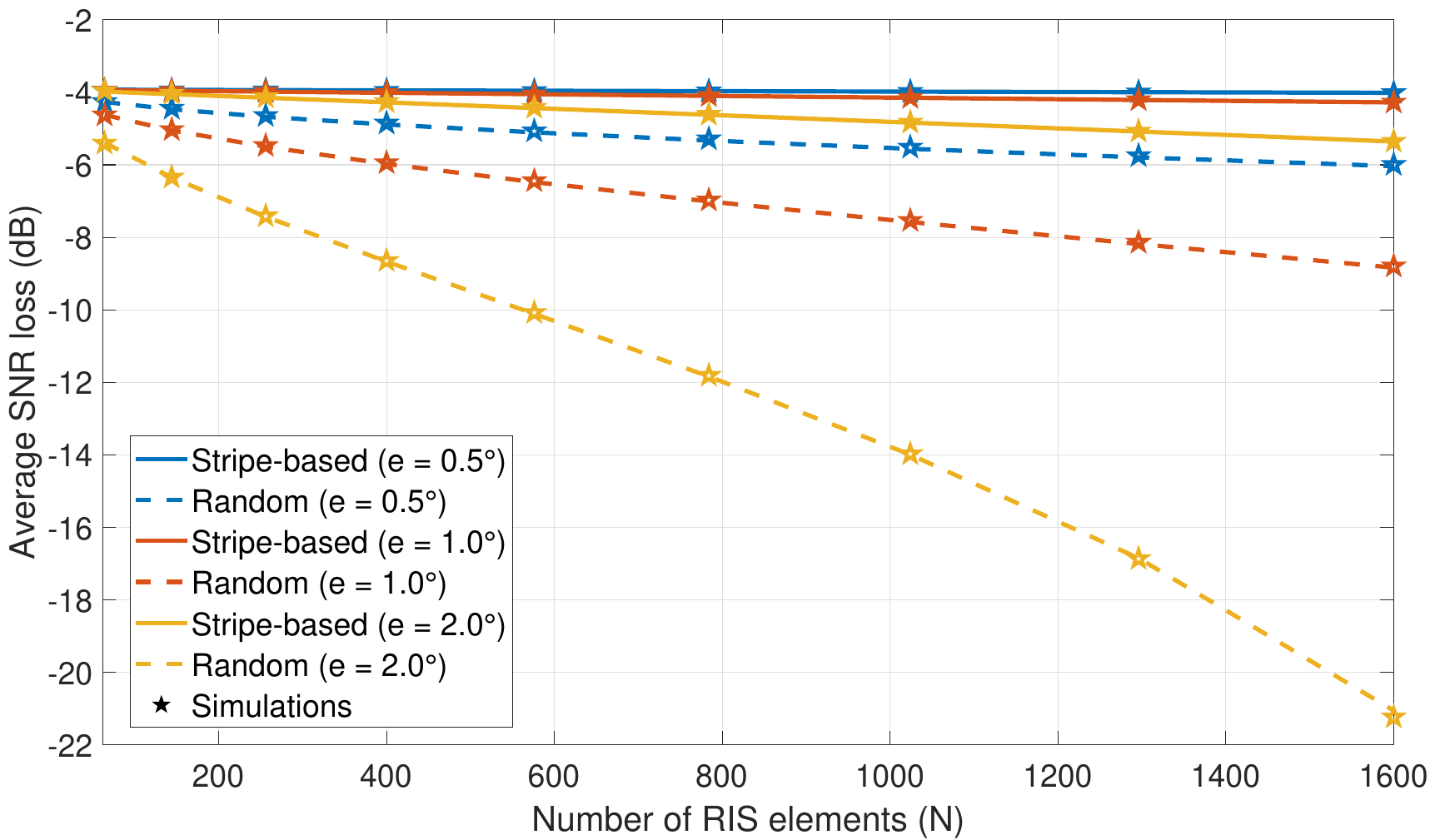}
    \caption{The analytical and simulation results for the average SNR loss values under various phase estimation errors using random bit flipping and stripe-based methods.}
    \label{fig:SNRLoss_RISSize}
\end{figure}

 The impact of phase-gradient mismatch on the SNR of a 1-bit RIS is evaluated in Fig.~\ref{fig:SNRLoss_RISSize}. The RIS configuration is parameterized by $(\Delta_x,\Delta_y)$, and impairments are modeled as a constant gradient error that affects both directions identically for simplicity, where $e=e_x=e_y$. Square RIS sizes $N=N_xN_y$ are swept with $N_x=N_y\in\{8,12,16,20,24,28,32,36,40\}$ and consider $e\in\{0.5^\circ,1^\circ,2^\circ\}$. For each $(N,e)$ pair, the SNR loss is computed relative to the ideal continuous-phase RIS gain and averaged over a dense uniformly distributed grid of far-field angle pairs $(\vartheta,\varphi)\in[0,360^\circ)$ to obtain geometry-averaged statistics. In addition to the structured bit-flip pattern induced by the gradient error, a benchmark is included in which the same overall fraction of incorrectly flipped RIS elements is applied through independent random flips.

The results show that structured gradient mismatch is significantly less destructive than random bit flips with the same flip rate. Across all RIS sizes, the structured case remains close to the inherent 1-bit quantization loss and exhibits only a mild additional degradation as $N$ increases, even for $e=2^\circ$. In contrast, the Random benchmark experiences a pronounced loss that grows with both RIS size and error magnitude, reflecting the rapid loss of coherent combining under uncorrelated flips. The analytical predictions closely match the simulation results across all tested RIS sizes and error levels, which confirms the accuracy of the proposed degradation models. In particular, the results show that SNR loss is not determined solely by the number of incorrectly flipped RIS elements, but also by how these flips are distributed across the aperture. Structured flip patterns lead to much smaller degradation than uncorrelated random flips with the same overall fraction.

\subsection{UAV Tracking Performance}
\begin{figure}[t]
    \centering
    \includegraphics[width=\linewidth]{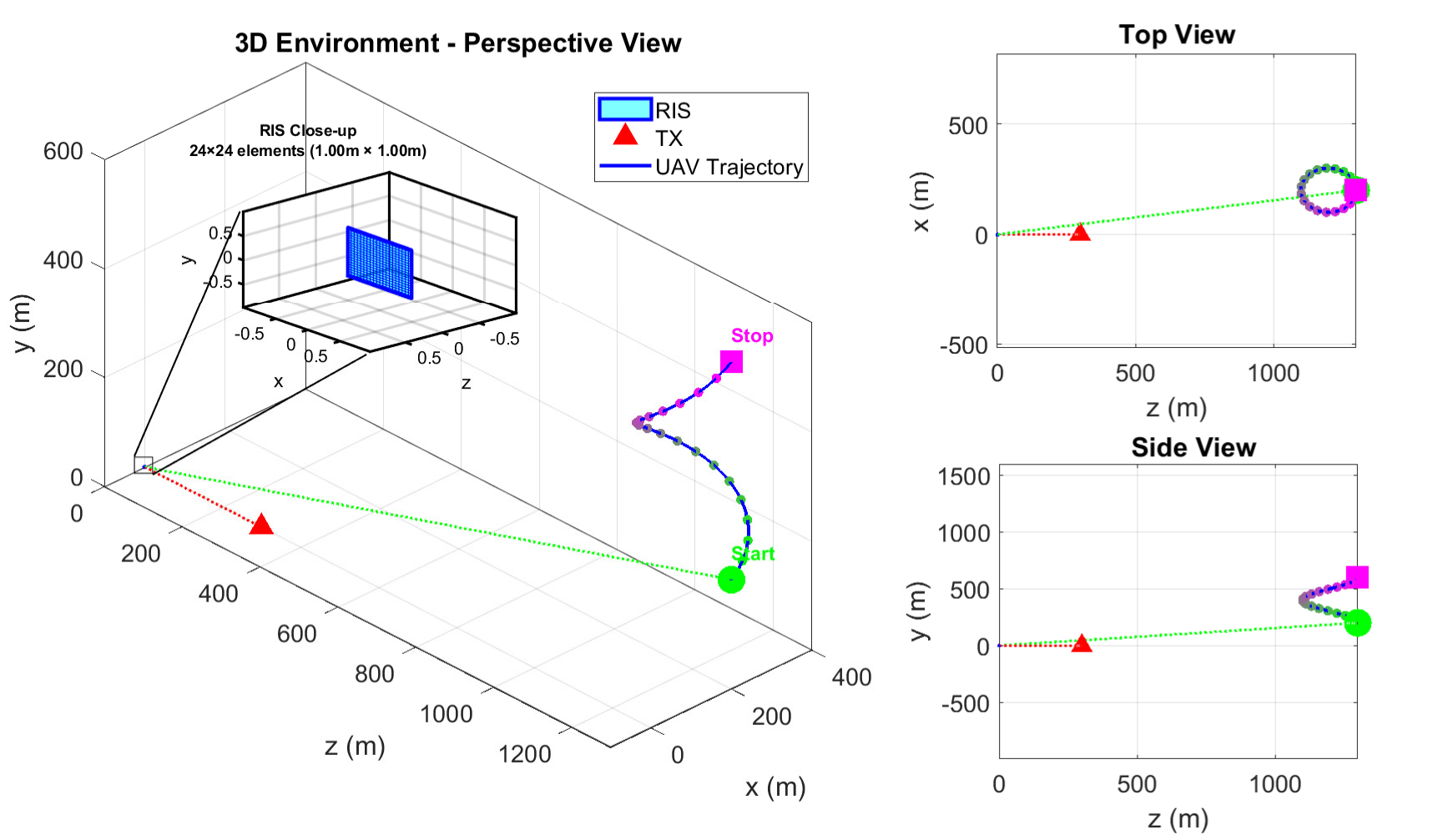}
    \caption{3D UAV trajectory.}
    \label{fig:UAV_Path}
\end{figure}

\begin{figure}[t]
    \centering
    \includegraphics[width=\linewidth]{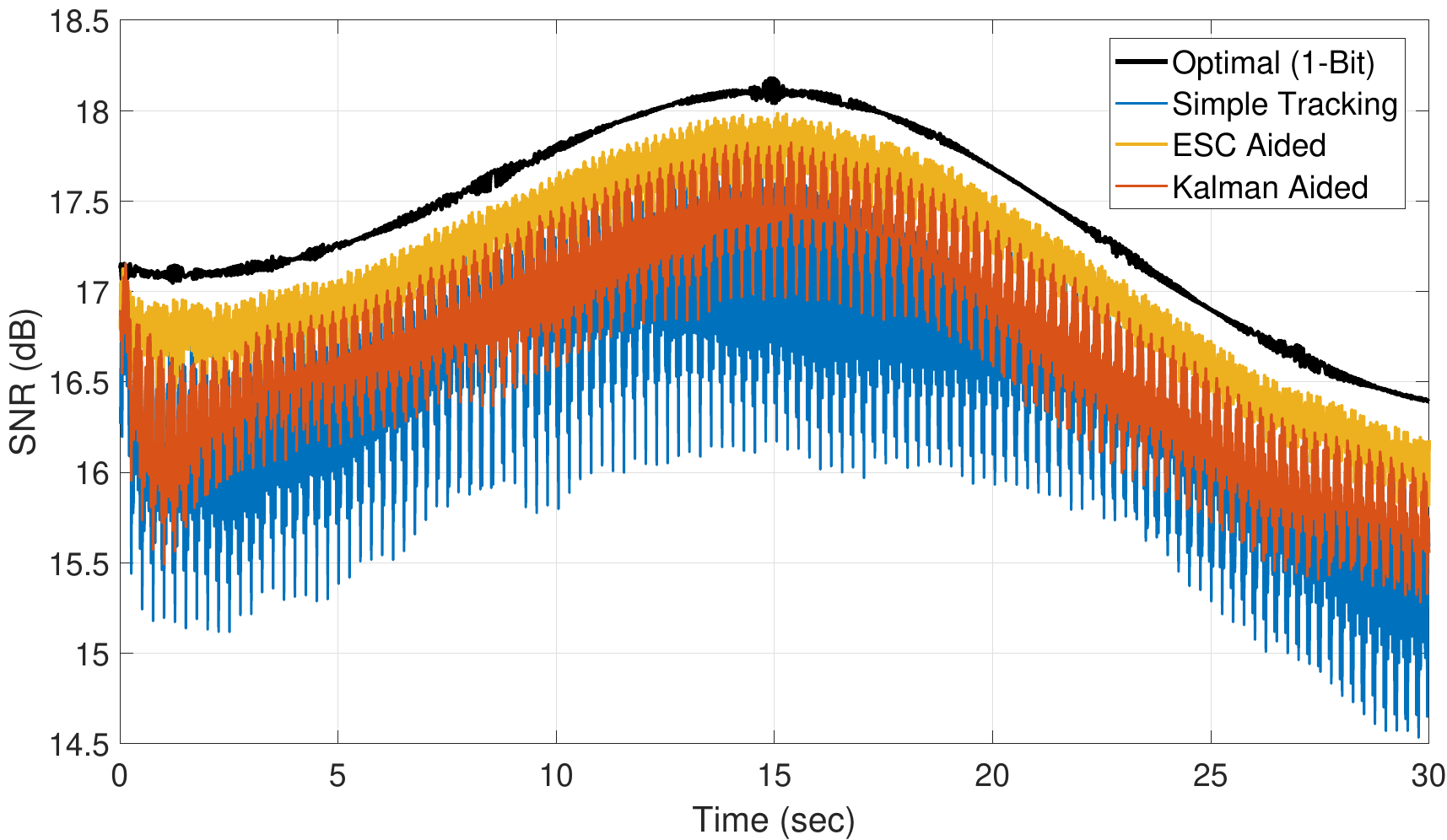}
    \caption{The averege received SNR during the UAV trajectory for different tracking algorithms.}
    \label{fig:Optimal_Stripe_Kalman_ESC}
\end{figure}

The RIS tracking results for a moving UAV receiver using stripe-based phase-gradient control are shown in Figs.~\ref{fig:UAV_Path}-\ref{fig:Optimal_Stripe_Kalman_ESC}. The transmitter is fixed at $[300,0,0]~\mathrm{m}$ with a transmit power of $70$ dBm. The receiver noise power is $-95$ dBm. The UAV follows the three-dimensional trajectory, which is taken from \cite{5717652} shown in Fig.~\ref{fig:UAV_Path} over a $30$ s interval. Received power is measured every $\delta_m=10$ ms. As references, the practically optimal 1-bit RIS configuration obtained via exhaustive per-element phase selection at each time instant is included.

All tracking methods operate on the stripe phase-gradients $(\Delta_x,\Delta_y)$. The recursive simple tracker uses a fine grid of resolution $N_f=5$ along each dimension, resulting in $N_f^2$ measurements per update cycle and an update duration $\delta_u = N_f^2 \delta_m$. The local search window width is selected proportional to $(V_{\max}/R_{\min})\delta_u$ with an additional safety margin. The Kalman-assisted tracker uses the same grid resolution $N_f=5$, but centers the search window around a predicted phase-gradient parameters state obtained from a constant-velocity Kalman filter with state vector $[\Delta_x,\Delta_y,\dot{\Delta}_x,\dot{\Delta}_y]^\mathrm{T}$. The extremum seeking controller applies sinusoidal dithers with amplitudes $a_x=a_y=1.5^\circ$ and frequencies $(\omega_x,\omega_y)=(35,28)$ Hz to the phase-gradient parameters, uses an averaging window of $N_a=7$ samples for gradient estimation, and updates the parameters with adaptation gain $\kappa=0.4$.

Fig.~\ref{fig:Optimal_Stripe_Kalman_ESC} presents the received SNR as a function of time. All three methods can track the moving UAV along its trajectory; however, they exhibit different SNR gaps relative to the practically optimal 1-bit configuration. The recursive simple search shows the largest ripple and the highest average loss (approximately $0.89$ dB). This is because each update sequentially scans multiple grid points, while the optimal phase-gradient pair continues to vary during the scan. As a result, the applied configuration becomes partially outdated before the update is completed. The Kalman-assisted method lowers the average loss (approximately $0.66$ dB) by predicting the parameter evolution and restricting the search region. This reduces the effective update duration and limits the mismatch caused by motion. The extremum seeking controller achieves the smallest average SNR loss (approximately $0.31$ dB). Instead of performing discrete grid sweeps, it applies continuous, low-amplitude perturbations, which reduces update latency and keeps the operation closer to the instantaneous SNR maximum. Overall, approaches that shorten the effective search interval or avoid explicit grid scanning yield more stable SNR behavior and smaller tracking error under receiver motion.

\begin{figure}[t]
    \centering
    \includegraphics[width=1\linewidth]{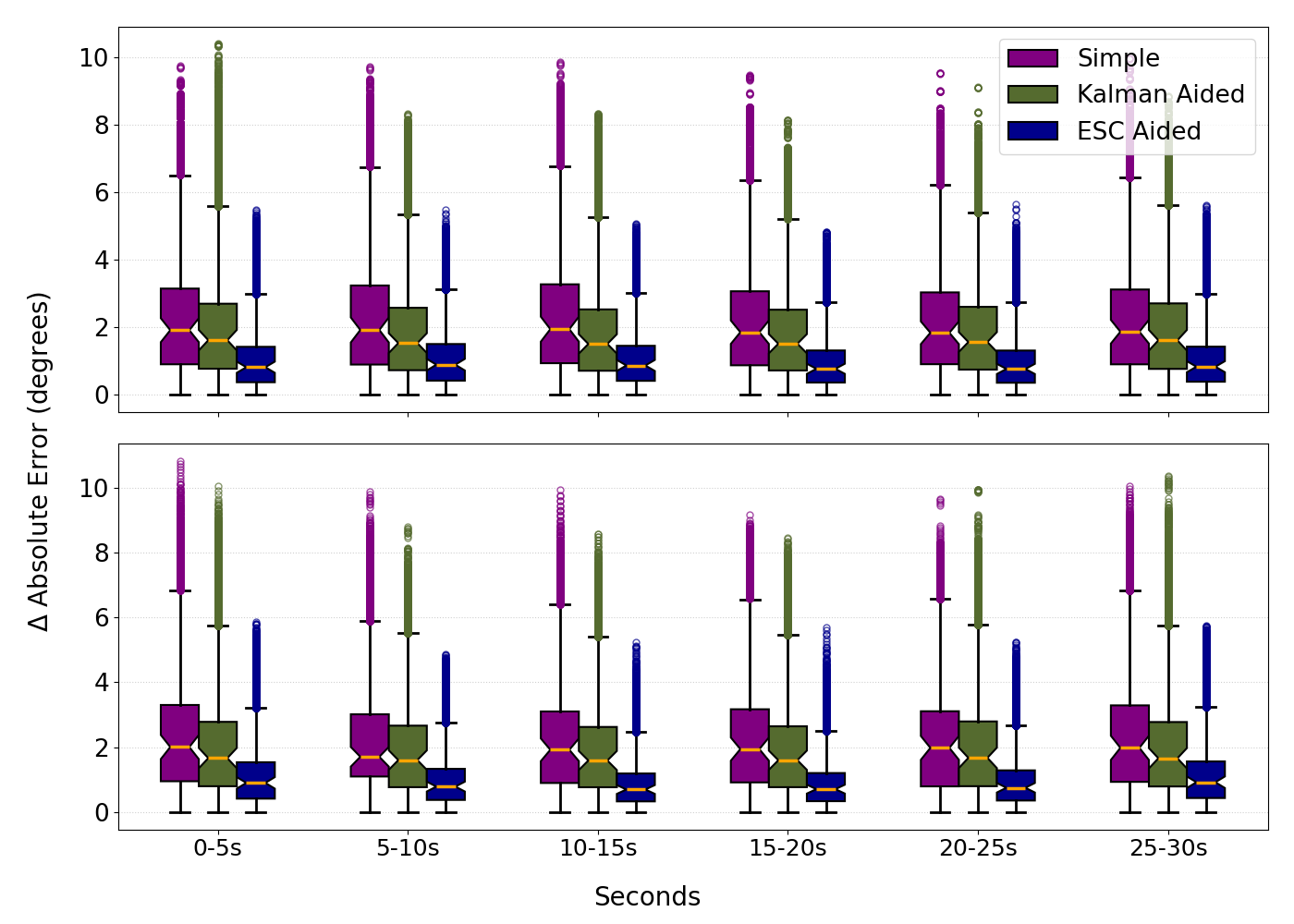}
    \caption{The absolute error boxplots for the combined direction parameters ($\Delta_x$ and $\Delta_y$).}
    \label{fig:delta_errors}\vspace{-10pt}
\end{figure}

\begin{figure}[t]
    \centering
    \includegraphics[width=1\linewidth]{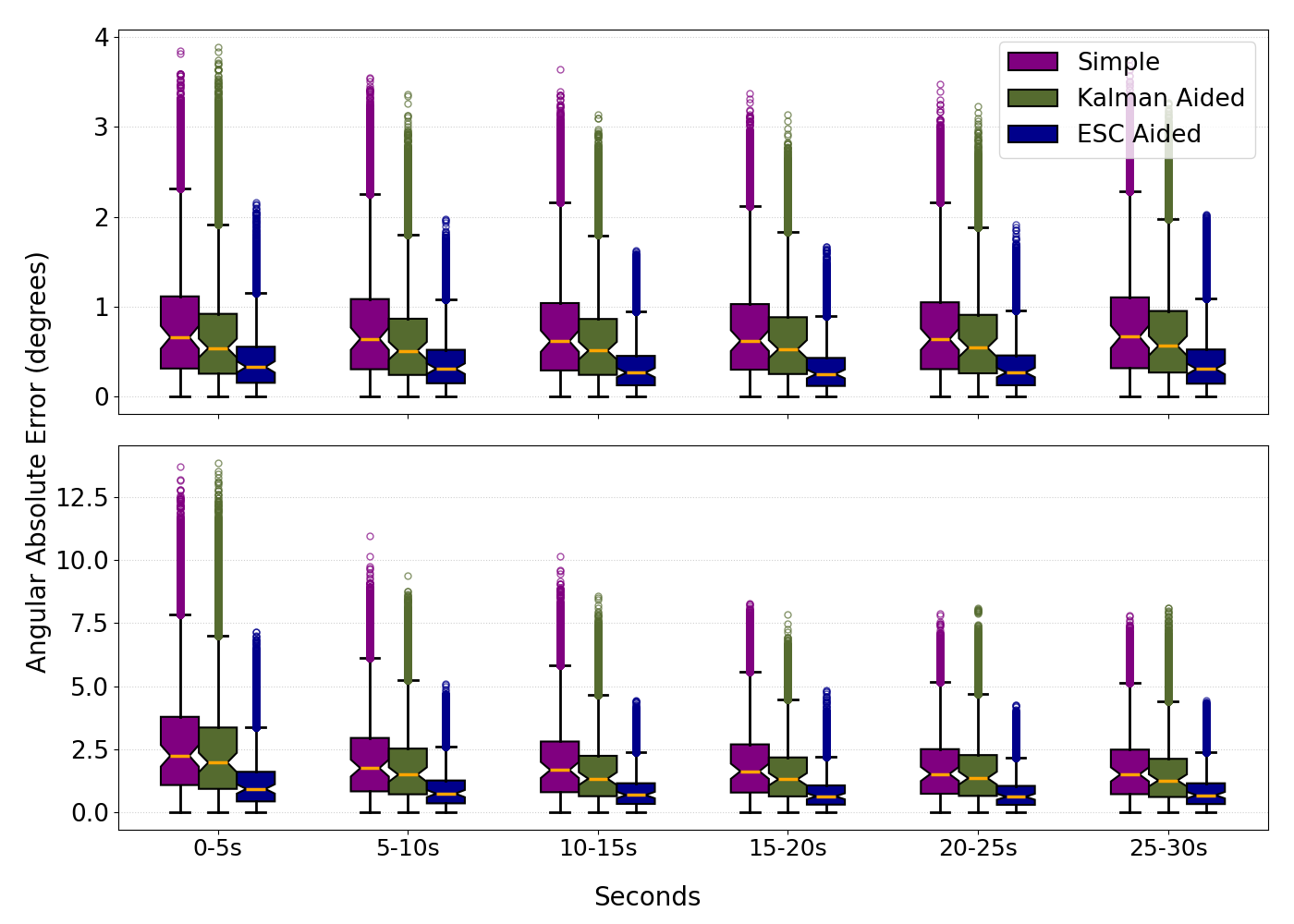}
    \caption{The absolute error boxplots for elevation and azimuth estimation results.}
    \label{fig:angular_errors}
\end{figure}

Figs.~\ref{fig:delta_errors} and~\ref{fig:angular_errors} provide a distribution-level assessment of the tracking accuracy, complementing the SNR comparison in Fig.~\ref{fig:Optimal_Stripe_Kalman_ESC}. For each method (Simple, Kalman Aided, and ESC Aided), we compute the absolute estimation errors with respect to the ground-truth values obtained from the geometry, namely $|\widehat{\Delta}_x-\Delta_x|$ and $|\widehat{\Delta}_y-\Delta_y|$ in Fig.~\ref{fig:delta_errors}, and the corresponding angular errors $|\widehat{\vartheta}-\vartheta|$ (elevation) and $|\widehat{\varphi}-\varphi|$ (azimuth) in Fig.~\ref{fig:angular_errors}. To obtain a compact time-resolved summary, the \SI{30}{s} flight interval is divided into six equal \SI{5}{s} segments, and for each segment, all samples across all time instants and Monte Carlo runs are pooled to form one box per method.

In each boxplot, the center orange line denotes the median error, the box edges indicate the first and third quartiles (IQR), whiskers extend to the most extreme non-outlier samples, and individual markers represent outliers. Hence, smaller medians and narrower IQRs imply more accurate and consistent tracking. Finally, these angular-error boxplots directly support the UAV localization discussion in Sec.~IV. Since the UAV direction $(\vartheta^{\mathrm{UAV}}(t),\varphi^{\mathrm{UAV}}(t))$ is inferred from the estimated combined direction parameters via \eqref{eqn:dx_loc}-\eqref{eqn:dy_loc}, the dispersion of $|\widehat{\vartheta}-\vartheta|$ and $|\widehat{\varphi}-\varphi|$ in Fig.~\ref{fig:angular_errors} quantifies the implicit localization accuracy achieved as a byproduct of RIS optimization.

\subsection{Experimental Validation of Stripe-based Optimization}
\begin{figure}[t]
    \centering
    \includegraphics[width=1\linewidth]{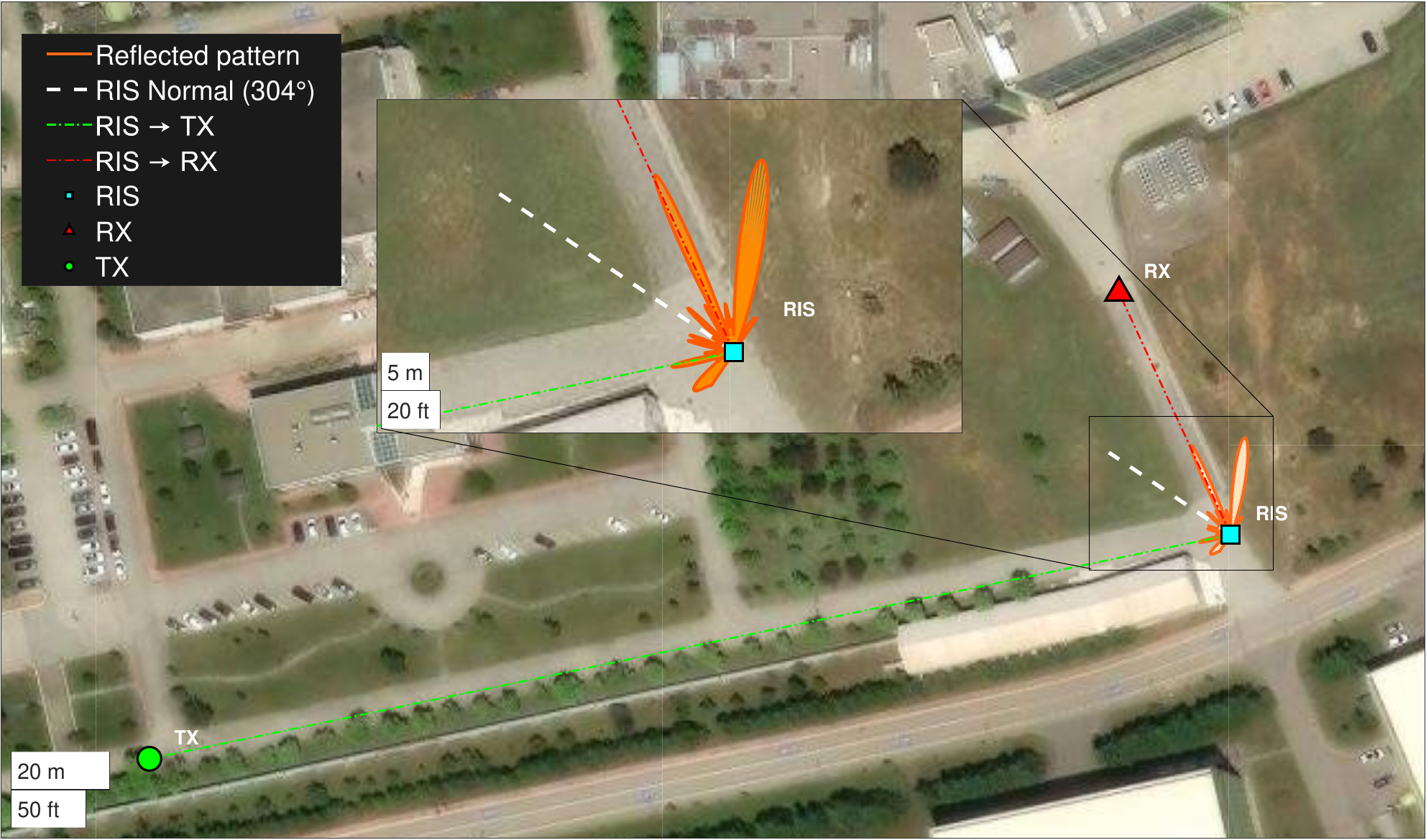}
    \caption{The measurement environment and the simulated RIS reflected radiation pattern.}
    \label{fig:Measurement_Map}
\end{figure}

The stripe-based optimization achieves high efficiency under two key assumptions, namely that both the BS and the UAV are located in the far field of the RIS with no line-of-sight path between them and that the propagation environment contains no strong scatterers that would generate significant multipath components from diverse angles. To validate the underlying assumptions and the practical viability of the proposed algorithm, empirical testing in a real-world environment is essential. Due to the operational complexity and deployment overhead associated with a UAV-integrated measurement setup, the experiments are performed outdoors at the TUBITAK Gebze campus, as illustrated in Fig. \ref{fig:Measurement_Map}, using a stationary transmitter–receiver pair located on the ground. An outdoor campus environment is chosen to meet the algorithm's far-field requirements.
ADALM-Pluto SDR transceivers are employed at both the transmitter (Tx) and receiver (Rx) nodes. During the measurements, the TUBITAK BILGEM-manufactured RIS \cite{11235096} operating at 3.75~GHz is remotely controlled through a Wi-Fi interface by a workstation. The RIS consists of $8 \times 8$ elements per tile, and a total of $3 \times 3$ tiles (i.e., $9$ tiles) are activated during the experiments, corresponding to an effective array size of $24 \times 24$ elements. For a half-wavelength element spacing, the resulting RIS aperture diagonal is approximately $D = 1.35$~m, yielding a Fraunhofer far-field distance of about $44$~m. During the experiments, the Rx is placed at a fixed distance of $54$~m from the RIS. The measurements are carried out for two different Tx locations, positioned at distances of $47$~m and $225$~m from the RIS, respectively. Therefore, the far-field requirements are satisfied for all Tx and Rx locations. Both the Tx and Rx are equipped with directive horn antennas with a half-power beamwidth of $40^\circ$. The antennas are oriented toward the RIS in order to suppress the direct Tx–Rx link and ensure that the dominant propagation path is RIS-assisted. For each measurement configuration, three RIS optimization strategies are evaluated: the $1 \times 1$ iterative method, the $4 \times 4$ iterative method, and the proposed stripe-based optimization algorithm. In addition, a baseline measurement is recorded with the RIS set to the \emph{off} state to quantify the relative gain achieved by each optimization approach. Due to the analog-to-digital converter (ADC) resolution of the SDR platform, the sampled complex baseband samples are depicted by integers in the range of $(-2047,\,2048]$. Thus, the average power of the sampled received signal can be calculated in decibels relative to full scale (dBFS) as \cite{10278759}
\begin{equation}
    P_{\mathrm{dBFS}} = 10 \log_{10} \left( \frac{1}{K} \sum_{k=1}^{K} |y[k]|^2 \right),
\end{equation}
where $y[k]$ denotes the complex baseband samples and $K$ is the number of samples used in the averaging window.

\begin{figure}[t]
    \centering
    \includegraphics[width=1\linewidth]{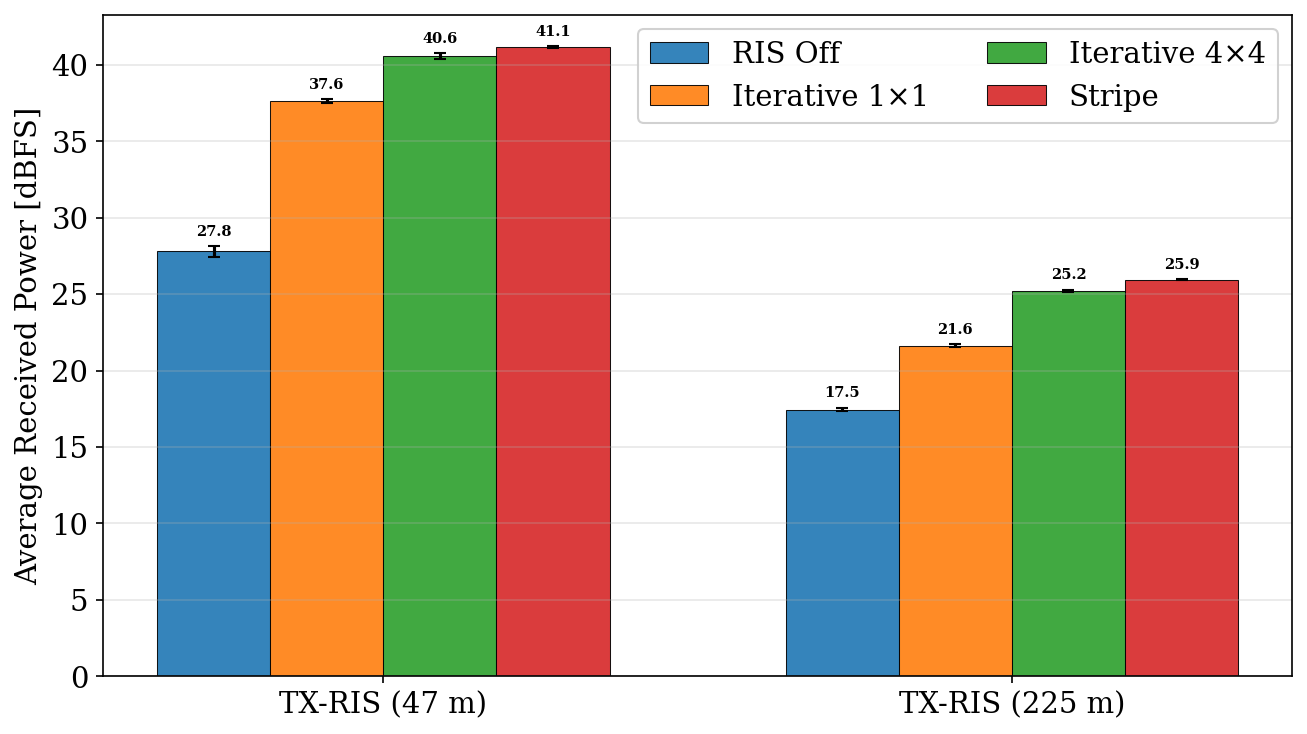}
    \caption{The measurement results of the average received power for different RIS optimization methods at two different Tx positions}
    \label{fig:Measurement_Method_Comparison_fixColor}
\end{figure}

\begin{figure}[t]
    \centering
    \includegraphics[width=1\linewidth]{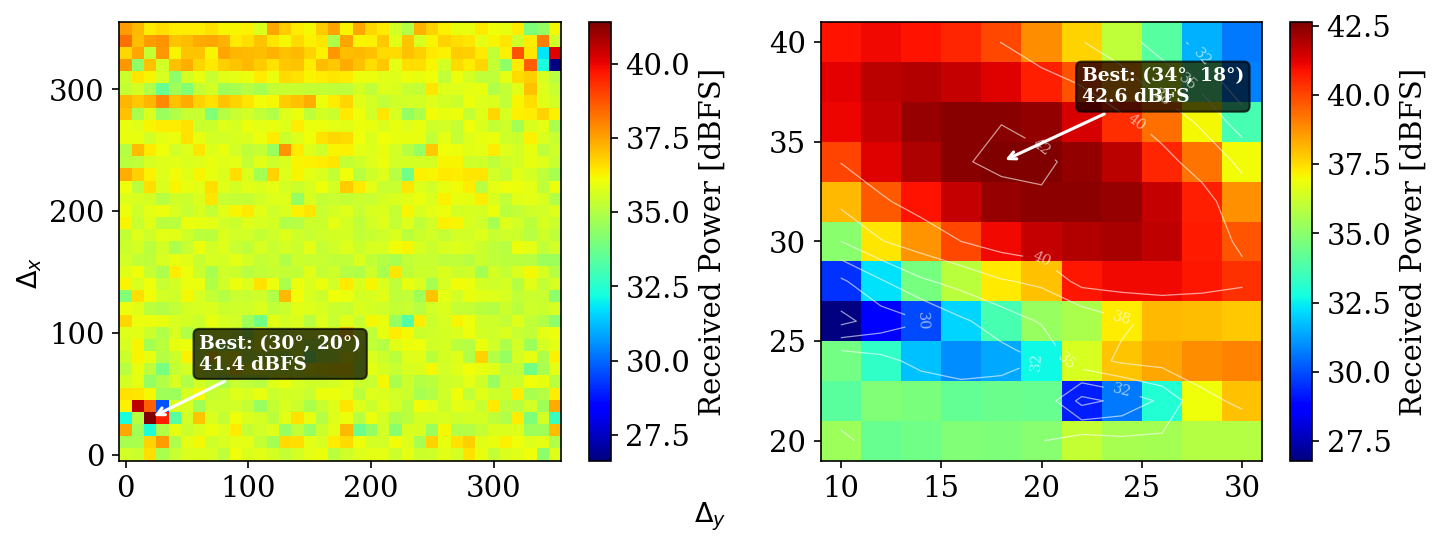}
    \caption{The measurement results of the instantaneous received power using the stripe-based method. The coarse stage is shown on the left, while the fine stage is on the right.}
    \label{fig:Measurement_Heatmaps}
\end{figure}

The measurement results of the average received power for different RIS optimization methods at two different Tx positions are shown in Fig.~\ref{fig:Measurement_Method_Comparison_fixColor}. The stripe-based method with $N_c = 36$ and $N_f = 20$ achieves the highest received power at both positions compared to the iterative algorithms. The measured instantaneous received power values obtained during the coarse and fine searches are depicted in Fig.~\ref{fig:Measurement_Heatmaps}. For the measurement setup with a Tx–RIS separation of 225~m, the coarse search is performed with a $10^\circ$ angular resolution, whereas the fine search uses a $1^\circ$ resolution. The optimum coarse-search parameters are measured as $(\Delta_x,\, \Delta_y) = (30,\,20)$, as shown in the left heatmap. The optimum fine-search parameters are $(34,\,18)$, as indicated in the right heatmap. A MATLAB simulation model of the measurement site is developed by placing the Tx, Rx, and RIS at the same locations as in the experiments. The optimal fine-search parameters $(34,\,18)$ are fed into the simulation environment to generate the RIS configuration. As shown in Fig.~\ref{fig:Measurement_Map}, the resulting radiation pattern verifies that the RIS directs the reflected signal toward the Rx, with an additional secondary sidelobe component.

\section{Conclusion}
In this study, a stripe-based RIS phase shift optimization method is presented to enable reliable UAV communication with a ground base station in LAWNs. The proposed approach also successfully tracks a mobile UAV, enabling passive sensing under realistic 3D movement of the UAV. The proposed low-complexity optimization method exploits the RIS’s structural phase-gradient to efficiently calculate its discrete phase adjustments while the UAV is in motion. Furthermore, by integrating a realistic 3D UAV movement model, two mobility-aware tracking schemes are developed and validated: a Kalman-aided predictor for smooth trajectory estimation and an ESC-aided approach for adaptive convergence under stochastic perturbations. The analysis and simulation results demonstrate that the proposed framework for an RIS-assisted UAV-based LAWN outperforms conventional benchmarks in convergence speed and computational efficiency, while maintaining robust, high-SNR connectivity even in the presence of phase estimation errors and low-SNR regimes. In addition, outdoor measurement experiments conducted with a real RIS prototype on a campus environment validate the practical feasibility of the proposed approach.

Future work will extend this framework to multi-UAV scenarios, investigating the mitigation of inter-user interference through coordinated multi-RIS reflection patterns. Additionally, the proposed tracking algorithms are validated on an experimental LAWN testbed to further assess real-world latencies and hardware impairments.

\bibliographystyle{IEEEtran}
\bibliography{main.bib}

\vfill
\end{document}